\documentclass[12pt]{iopart}

\usepackage{graphicx}

\begin{document}


\title{Generic modes of consensus formation in stochastic language dynamics}

\author{R.\ A.\ Blythe} \address{SUPA, School of Physics and Astronomy, University
of Edinburgh, Mayfield Road, Edinburgh EH9 3JZ}

\begin{abstract}
We introduce a class of stochastic models for the dynamics of two linguistic variants that are competing to become the single, shared convention within an unstructured community of speakers. Different instances of the model are distinguished by the way agents handle variability in the language (i.e., multiple forms for the same meaning). The class of models includes as special cases two previously-studied models of language dynamics, the Naming Game, in which agents tend to standardise on variants they have encountered most frequently, and the Utterance Selection Model, in which agents tend to preserve variability by uniform sampling of a pool of utterances. We reduce the full complexities of the dynamics to a single-coordinate stochastic model which allows the probability and time taken for speakers to reach consensus on a single variant to be calculated for large communities. This analysis suggests that in the broad class of models considered, consensus is formed in one of three generic ways, according to whether agents tend to eliminate, accentuate or sample neutrally the variability in the language. These different regimes are observed in simulations of the full dynamics, and for which the simplified model in some cases makes good quantitative predictions. We use these results, along with comparisons with related models, to conjecture the likely behaviour of more general models, and further make use of empirical data to argue that in reality, biases away from neutral sampling behaviour are likely to be small.
\end{abstract}

\maketitle


\section{Introduction}

Statistical mechanical modelling is increasingly being used as a methodology to reproduce and predict large-scale, emergent regularities in human systems. Recent examples where good quantitative agreement with empirical data has been observed (sometimes reported under the guise of `agent-based' modelling, which is essentially the same approach) include the flow properties of highway traffic \cite{kno04}, some aspects of stampedes arising from crowd panic \cite{hel07} and various distributions relating to firms in an economy \cite{del07}. The similarity with traditional condensed matter applications is that macroscopic properties are obtained by averaging over an ensemble of microscopic degrees of freedom. The key difference, however, is that the nature of the underlying microscopic interactions is itself a source of uncertainty. Here one hopes to be saved by some kind of `universality principle', which we will loosely interpret as meaning that a wide class of systems differing in microscopic details nevertheless display similar generic properties at large scales. In this work, we scrutinise the relevance of this idea in the context of \emph{social dynamics}. This is a relatively new, but nevertheless burgeoning application domain in statistical mechanics (see \cite{cas07} for a comprehensive review) that promises to contribute to the general quantitative understanding of cultural origins, evolution and change \cite{cav81,boy05}.

Specifically, we focus on the situation where speakers of a language have a choice of two different ways of saying the same thing. Which of these two \emph{variants} is uttered by a speaker at a given time is assumed to be a function of that speaker's exposure to utterances produced by herself\footnote{In common with earlier works, we will use male and female gender pronouns when referring to listening and speaking agents, respectively.} and other members of her community at earlier times. (This is in the spirit of the \emph{usage-based} approach to linguistics \cite{pie03}). Over time, the relative frequencies with which specific variants are used may fluctuate, and may perhaps reach a steady state in which both variants are used with some non-zero frequency, or one variant may go extinct. A lot of theoretical attention has been devoted to the latter case, and has been described as \emph{conventionalisation} in a linguistic context \cite{cro00}, \emph{consensus} in opinion dynamics \cite{cas07} (in which agents hold one of a number of variant \emph{opinions} whose frequencies change over time) and \emph{fixation} in population genetics \cite{ewe04}. Meanwhile, this process has also been of empirical interest, for example, in sociolinguistic studies charting the rise of one set of phonetic realisations of vowel sounds over another in such locations as Philadelphia \cite{lab01} and New Zealand \cite{gor04}, or in the adoption of an innovative technology such as hybrid corn in Iowa \cite{rog03}.

Even within this fairly restricted context of consensus formation (the terminology that we will adopt here), a wide variety of models have been proposed \cite{cas07}. How this space of models is structured is at present unclear. In this work we seek to gain some understanding of what this structure might be by introducing and analysing a class of models that interpolates between two distinct types of individual agent (speaker) behaviour and includes as specific instances contrasting models that have been previously discussed with reference to language dynamics. One feature that all these models have in common is \emph{neutrality}: one phonetic realisation of the vowel appearing in the word `trap', for example, is not assumed to be better suited to the task than any other, and hence \textit{a priori} preferred by all speakers. Where the models differ is in the rule used to decide which variant to utter in an interaction. Two generic strategies, namely \emph{sampling} and \emph{maximising}, are encapsulated. A sampler produces a variant with a probability equal to the frequency that she perceives it to be used in her community. A maximiser, on the other hand, chooses the variant she believes to be used \emph{most} frequently in the community. In model systems, sampling behaviour is represented by the Voter Model \cite{cli73}, the Utterance Selection Model of language change \cite{bax06} and relatives. Meanwhile, maximising can be identified in nonlinear Voter Models \cite{oli93,dro99,dor01,sch08} and the Naming Game \cite{bar06}. If interpreted as a tendency to preserve or eliminate grammatical irregularity in an artificial language, these two behaviours have also been identified in children and adults respectively during psycholinguistics experiments \cite{hud05} and further shown to affect the emergent structure of a language acquired by successive generations of speakers \cite{kir07}.

In this work, we establish the generic modes of consensus formation exhibited by populations of interacting speakers that are differentiated through the behaviour of the individual agents that they are composed of. Our strategy is to start with two concrete models, namely versions of the Voter / Utterance Selection Model and Naming Game, in which agents invoke local sampling and maximising rules respectively. In Section~\ref{rules} we recall the definitions of these models so as to establish that, when restricted to two variants, their microscopic update rules differ in two fundamental ways. One involves a bias towards categorical use of a single variant that is imposed by the listener in any given interaction; the other a similar bias applied by the the speaker. By varying the strength of these biases, a two-parameter \emph{hybrid model} is generated within which we find only three distinct modes of consensus formation.  The modelling approach is similar to that followed in \cite{bar07}, in which a single-parameter generalisation of the Naming Game is constructed by employing one of the microscopic dynamical rules stochastically. Despite the fact that the dynamics of this latter generalisation cannot be reproduced by special choices of the parameters in the present hybrid model, we find that the same generic phase behaviour is common to both families of models, thereby adding weight to the hypothesis that the collective behaviour is only weakly affected by changes in the microscopic dynamics (except near a transition point).

The model's phase diagram we establish initially by examining deterministic equations of motion, presented in Section~\ref{det}. We find two distinct regimes, one in which consensus on the majority variant is reached, and another in which both variants coexist in perpetuity, separated by a line in the parameter space in which the variant frequencies do not change over time. The addition of noise allows consensus to be reached for all parameter combinations, and it is these stochastic effects that are of greatest interest in the context of consensus formation. In particular, we seek to calculate the probability consensus on a particular variant will be reached, and the time taken to do so, given the initial condition. Since the full stochastic equations of motion are rather complicated, we use observations on the nature of the deterministic trajectories to suggest a means to reduce to a single stochastic dynamical variable. These simplified dynamics are formulated in Section~\ref{stoch} and analysed in Section~\ref{large} for large communities, from which we find evidence that the phase diagram is robust to noise. Furthermore, we find that the typical consensus time in a maximising community is of order $N\ln N$ interactions (where $N$ is the community size), of a pure sampling community of order $N^2$ interactions, and exponentially large in $N$ when speakers exhibit an `anti-maximising' behaviour, i.e., a tendency to prolong variability in the language. The validity of the simplified dynamics as a proxy for the full dynamics is explored through computer simulation in Section~\ref{sim}. We find that its predictions for the probability that a particular variant wins out agree rather well with simulation, and those for the mean time to do so are qualitatively correct but in some regimes display some quantitative differences from the values observed in simulation.

Whilst these consensus statistics are by now well established for the Voter Model and its relatives \cite{soo05,soo08,bax06,bax08a}, only the deterministic behaviour of the Naming Game in the two-variant regime has been treated analytically in previous works \cite{cas07,bar07,bar08}. Our findings for the \emph{dynamics} of the relaxation to the state of consensus in the presence of noise therefore complement existing studies of the static critical phenomena seen in family of models that include the Voter and Ising Models as special cases \cite{oli93,dro99,dor01}. In the concluding section, we return to the question of universality and make conjectures for the likely consensus properties for more general models of language dynamics, based on the insight gained from the hybrid model within the simplified single-coordinate approach.  Finally, we confront the hybrid model with empirical data for new-dialect formation \cite{gor04,tru04} to demonstrate the possibility that, in a real human system, a behavioural bias away from pure sampling behaviour may in fact be quite small.

\section{The hybrid model}
\label{rules}

The family of models under consideration has a single community of $N$ agents evolving by a sequence of interactions between pairs of randomly-chosen individuals. One of each pair is designated with probability $\frac{1}{2}$ the \emph{speaker} and the other the \emph{listener}. Since all pairs of speakers interact equally often, this defines an \emph{unstructured} or \emph{mean-field} community. Although clearly real communities exhibit structure, this simplification provides a substrate upon which we can precisely identify fundamental similarities and differences between models, the main motivation behind this work.

The model language comprises two variant forms for a single meaning. For example, two different phonetic realisations of a vowel sound, or two different words for an object. We denote these variants $A$ and $B$, and call their uttered realisations \emph{tokens} of the respective variants. We refer to a speaker's internal representation of the frequency that $A$ and $B$ variants are used in the community as her \emph{store}. In general, samplers and maximisers use the information in the store in different ways to decide which variant to use, and respond differently to an uttered token. In the concrete implementations that have previously been proposed as models for language behaviour, the Utterance Selection Model and the Naming Game, we find that tokens are produced in the same way in both models, and the distinction between sampling and maximising is manifested after the production event. It is this distinction that will form the basis of the hybrid model, which we define after recalling the two distinct limiting cases.

\paragraph{Sampling behaviour: the Voter Model and its relatives}

\begin{figure}
\begin{center}
 \includegraphics[width=0.75\linewidth]{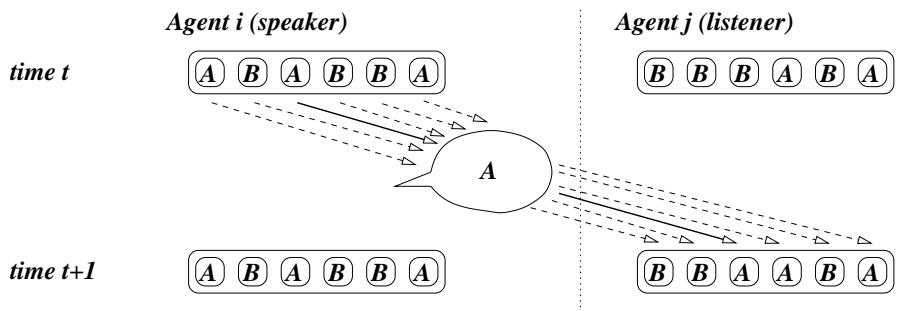}
\end{center}
 \caption{\label{sampler} Illustration of the update rules for the sampling-based Voter and Utterance Selection Models.  Each set of arrows corresponds to events that occur with equal likelihood; the production and replacement events realised in the illustrated interaction are indicated by solid lines.}
\end{figure}

A population of sampling agents can be implemented as follows. Each agent retains a fixed-sized store of previously heard tokens. When an agent is required to utter a token, she simply chooses one at random from the store. The listener meanwhile replaces one of his stored tokens, chosen at random, with a copy of the token produced by the speaker. As a consequence, each speaker maintains a sample of previously encountered tokens whose relative proportions of $A$s and $B$s will reflect that of the wider community, albeit subject to noise arising from the stochasticity in the choice of speaker-listener pairs, token production and the finite size of the store. These dynamics are illustrated in Fig.~\ref{sampler}.

If each speaker's store contains only a single token, this model corresponds exactly with the Voter Model on the complete graph (fully connected network) or the Moran model in population genetics \cite{ewe04,bly07}. If the stores are large, and both speaker and listener produce tokens in an interaction, and both retain copies of their own and their interlocutor's utterances, we recover the Utterance Selection Model \cite{bax06}. It turns out the consensus-formation behaviour of the model is essentially the same, no matter the size of the store or whether the roles of speaker and listener are separate or conflated \cite{bly07,bax08a}. In a finite-sized community, one variant will eventually go extinct; the probability variant $A$ wins out is equal to its initial frequency in a mean-field community; and the distribution of extinction times is the same for all models once time has been rescaled by a factor that depends on the size of the stores and the precise manner of the speaker-listener interaction (as long as the essential sampling behaviour for production and replacement of tokens is retained).

\paragraph{Maximising behaviour: the Naming Game}

\begin{figure}
\begin{center}
 \includegraphics[width=0.85\linewidth]{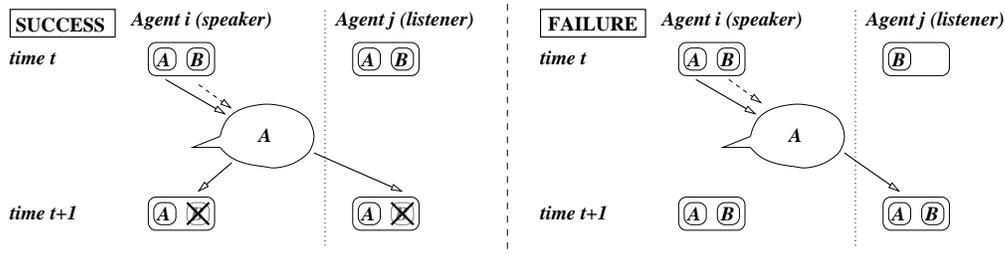}
\end{center}
 \caption{\label{naminggamer} Illustration of the update rules for the maximising behaviour as implemented in the Naming Game when restricted to two variants. The two types of interaction, success and failure, are shown. After a successful interaction, both listener and speaker delete from their stores any instances of tokens other than that just uttered; after a failure, the uttered token is added to the listener's store. Sets of dashed and solid arrows are to be interpreted as in Fig.~\ref{sampler}.}
\end{figure}

One way to implement a maximising behaviour is as formulated in the Naming Game \cite{bar06}. In this model, a speaker retains at most one token of a given variant in the store. Again, when placed in the speaking role, an agent selects one of the tokens from her store uniformly at random for production. In this model, maximising behaviour is implemented by the listener: if the token uttered by a speaker matches a token in the listener's store, the interaction is deemed a `success' and the listener erases all instances of other variants from the store. That is, the listener associates the locally maximal variant (among his store $+$ the uttered token) uniquely with the target meaning after a successful interaction. After a successful interaction, the association is further strengthened through the speaker also erasing any tokens in her store that do not represent the variant she has just uttered. If the interaction is a failure (the listener does not have a matching token in his store), his store is extended to include a token of the uttered variant. See Fig.~\ref{naminggamer}.

\paragraph{The hybrid model}

With just two variant forms, $A$ and $B$, in the language, speakers in the Naming Game behave in one of three ways: produce only $A$; produce only $B$; or produce $A$ or $B$ with equal probability. These same three production rules are realised in the Voter Model if each speaker's store holds two tokens. We thus denote these three states as $AA$, $BB$ and $AB$ respectively, and further refer to the $AA$ and $BB$ states as \emph{consistent} (since speakers consistently produce a single variant), and the $AB$ state as \emph{inconsistent}. Consensus is reached when all agents are in the same consistent state. The same production rule, namely sampling one of the two variants in the store, will be applied by speakers in all instances of the hybrid model.

To place the differing behaviour of the listeners between the two models on the same footing, we reformulate the maximising rule that applies in the Naming Game as follows: if the listening agent is in the inconsistent state, he places a copy of the uttered token into his store, overwriting the token that doesn't match. In the Voter Model, this token is overwritten with probability $\frac{1}{2}$. These two rules can be unified by making this probability a variable parameter. The prescription in the hybrid model is to replace the non-matching token with probability $\frac{1}{2}(1+b)$, where $b$ is a maximisation bias parameter, which, if equal to $0$ recovers Voter-like behaviour, and if equal to $1$ recovers Naming-Game-like behaviour. Note that negative $b$, $b\ge-1$, is possible; then the listener exhibits an `anti-maximisation' behaviour, i.e., a reluctance to adopt consistent use of a single variant. When the listening agent is in a consistent state, the listener update for both the Voter Model and Naming Game is the same---viz, random replacement of one of the tokens in the store.

This leaves us with the speaker update rule of the Naming Game to implement. This occurs when the speaker's uttered token matches one in the listener's store: the `success' of the interaction is communicated to the speaker, who then adopts the corresponding consistent state. In the Voter Model, this never happens; in the Naming Game it takes place with probability $1$. The obvious way to unify these models is to apply this rule with a \emph{copy} probability $c$. That is, if the listener is in a consistent state at the end of the interaction, the speaker copies that state with probability $c$.

\begin{figure}
 \begin{center}\includegraphics[width=0.65\linewidth]{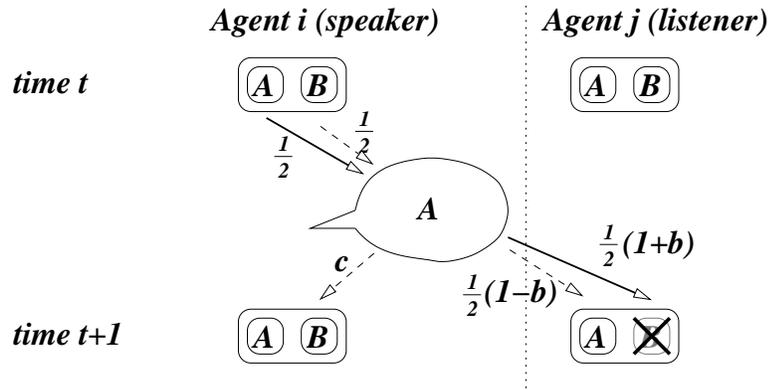}\end{center}
  \caption{\label{hrulefig} Illustration of an update rule for the hybrid model that includes the Voter / Utterance Selection Model and the two-variant Naming Game as the special cases $b=c=0$ and $b=c=1$ respectively.  Labels against the arrows indicate the probability the corresponding event (production or replacement) takes place.}
\end{figure}

Thus we arrive at a two-parameter family of models that interpolates between the pure sampling behaviour of the Voter / Utterance Selection Model and its relatives ($b=c=0$) and maximisation as implemented in the Naming Game ($b=c=1$). These dynamics are illustrated in Fig.~\ref{hrulefig}.  It is also convenient to summarise the  model definition through the set of transition probabilities presented in Table~\ref{rates}. In the table we also specify the changes in the number of agents $n_{\sigma}$ in each of the three states $\sigma\in\{AA,AB,BB\}$, and also the changes in the total number of tokens $m_{A}$ and $m_{B}$ across all stores in the community. In the following sections we will use this table to derive the deterministic and stochastic components of the dynamics.

\begin{table}
\begin{center}
\begin{tabular}{c|c|c||c|c|c||c|c}
$(\sigma,\lambda)$ & $(\sigma',\lambda')$ & $P(\sigma,\lambda \to \sigma',\lambda')$ & $\delta n_{AA}$ & $\delta n_{AB}$ & $\delta n_{BB}$ & $\delta m_A$ & $\delta m_B$ \\ \hline
$(AA,BB)$ & $(AA,AB)$ & 1 & 0 & 1 & -1 & 1 & -1 \\ \hline
$(BB,AA)$ & $(BB,AB)$ & 1 & -1 & 1 & 0 & -1 & 1 \\ \hline
$(AA,AB)$ & $(AA,AA)$ & $\frac{1}{2}(1+b)$ & 1 & -1 & 0 & 1 & -1 \\ \hline
$(BB,AB)$ & $(BB,BB)$ & $\frac{1}{2}(1+b)$ & 0 & -1 & 1 & -1 & 1 \\ \hline
& $(AA,AA)$ & $\frac{1}{2}c$ & 1 & -1 & 0 & 1 & -1 \\
\raisebox{1.5ex}[0pt]{$(AB,AA)$} & $(AB,AB)$ & $\frac{1}{2}$ & -1 & 1 & 0 & -1 & 1 \\ \hline
& $(BB,BB)$ & $\frac{1}{2}c$ & 0 & -1 & 1 & -1 & 1 \\
\raisebox{1.5ex}[0pt]{$(AB,BB)$} & $(AB,AB)$ & $\frac{1}{2}$ & 0 & 1 & -1 & 1 & -1 \\ \hline
& $(AA,AA)$ & $\frac{1}{4}(1+b)c$ & 2 & -2 & 0 & 2 & -2 \\
& $(AB,AA)$ & $\frac{1}{4}(1+b)(1-c)$ & 1 & -1 & 0 & 1 & -1 \\
\raisebox{1.5ex}[0pt]{$(AB,AB)$} & $(AB,BB)$ & $\frac{1}{4}(1+b)(1-c)$ & 0 & -1 & 1 & -1 & 1 \\
& $(BB,BB)$ & $\frac{1}{4}(1+b)c$ & 0 & -2 & 2 & -2 & 2 \\
\end{tabular}
\end{center}
\caption{\label{rates} Probability $P(\sigma,\lambda \to \sigma',\lambda')$  that a speaker-listener pair $(\sigma, \lambda)$ makes the transition to $(\sigma',\lambda')$ after having been chosen to interact. The columns headed $\delta n_{AA}$, $\delta n_{AB}$ and $\delta n_{BB}$ indicate changes in the number of $AA$, $AB$ and $BB$ agents as a result of the transition. Likewise, those headed $\delta m_{A}$ and $\delta m_B$ give the change in the total number of $A$ and $B$ tokens stored by all agents in the community.}
\end{table}

\section{Deterministic equations of motion}
\label{Sdet}

Deterministic equations of motion are obtained for the fraction of agents in state $\tau$, $x_{\tau} = n_{\tau}/N$, by summing over all possible changes in $n_{\tau}$ that can occur in an interaction, given the state of the system at time $t$ and weighted by the probability that the change occurs. In a mean field community, the probability that a randomly-chosen speaker and listener are in states $\sigma$ and $\lambda$ respectively is $x_{\sigma} x_{\lambda}$ (with a correction of order $1/N$ that we shall neglect). The probability that this pair then changes state to $\sigma', \lambda'$, and the corresponding changes in $n_{\tau}$ can then be read off from Table~\ref{rates}. By performing the sum, one finds
\begin{equation}
\label{det}
\delta x_{\tau} = \frac{1}{N} \sum_{\sigma,\lambda} x_{\sigma} x_{\lambda} \sum_{\sigma',\lambda'} P(\sigma,\lambda \to \sigma',\lambda') \delta n_{\tau}(\sigma,\lambda \to \sigma',\lambda') \;.
\end{equation}
In the large-$N$ limit, the changes $\delta x_{\tau}$ are small, and we can approximate the left-hand side of this equation as the time derivative $\dot{x}_{\tau}$.

The resulting equations of motion are most conveniently stated in the space of \emph{token frequencies}, $x_{A} = m_{A}/(2N)$ and $x_{B} = m_{B}/(2N) = 1-x_{A}$, within the aggregated store of the entire community (hence the factor of $2N$). The dynamics cannot be closed in terms of these two frequencies---one also needs to keep the fraction of inconsistent speakers $x_{AB}$ for a complete description. After some algebra, one finds
\begin{eqnarray}
\label{xA}
\dot{x}_{A}(t) &=& \frac{\mu}{2N} x_{AB}(x_A - x_B) \\
\label{xAB}
\dot{x}_{AB}(t) &=& \frac{1}{N} \left[ 2 x_A x_B - \left(1+\mu\right) x_{AB} - \frac{bc}{2} x_{AB}^2 \right] \\
\label{xB}
\dot{x}_{B}(t) &=& \frac{\mu}{2N} x_{AB}(x_B - x_A)
\end{eqnarray}
where we have introduced the key parameter
\begin{equation}
\mu = \frac{b+c}{2} \;,
\end{equation}
the arithmetic mean of the bias and copy parameters introduced to interpolate between the Voter / Utterance Selection Model and the Naming Game. One can show that with $b=c=1$, one recovers the expressions previously presented in the context of the Naming Game \cite{bar06,cas07} and a model for bilingualism \cite{cas06}.

Although noise has been neglected, these equations involve no further approximations. Therefore we may make the following observations.

If the parameter $\mu=0$, the mean frequencies of $A$ and $B$ tokens is conserved. This behaviour corresponds to that of the Voter Model and its relatives, within which it is known that consensus is brought about by a fluctuation that typically occurs after a time of order $N^2$ interactions \cite{kim69}. We anticipate then that this voter-like behaviour will be exhibited not just at the point $b=c=0$, but along the line $c=-b$ (recall that $b$ is permitted to be negative).

For non-zero $\mu$, we see that inconsistent agents present in the community induce an effective interaction between $A$ and $B$ token frequencies. If $\mu>0$, any difference in their number is amplified by the dynamics, whereas if $\mu<0$, a `restoring force' opposes these differences. This suggests consensus will be reached quickly when $\mu>0$, but will never be reached (in the absence of noise) when $\mu<0$. This allows us to draw a phase diagram for the space of models spanned by the parameters $b$ and $c$ where the line $\mu=0$ separates a phase in which consensus is reached rapidly from one in which the onset of consensus is delayed (to infinity, in the noise-free dynamics). See Figure~\ref{phasediag}. As we will see below, this phase diagram is retained when we consider a stochastic version of the dynamics.

\begin{figure}
\begin{center}
 \includegraphics[width=0.55\linewidth]{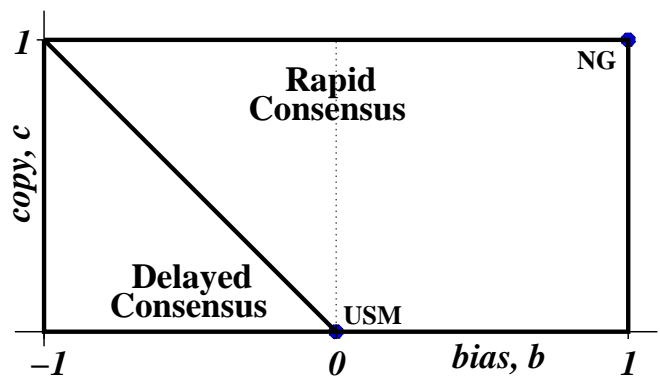}
\end{center}
\caption{\label{phasediag} Phase diagram for the hybrid model, as predicted by the deterministic equations of motion (\ref{xA})--(\ref{xB}).}
\end{figure}

The fixed-point structure of equations (\ref{xA})--(\ref{xB}) yields slightly more information about these phases. Two fixed points are at $(x_A,x_{AB},x_B)=(1,0,0)$ and $(0,0,1)$ and correspond to the state of consensus. These are stable when $\mu>0$. In the deterministic dynamics, the token that is initially in the majority always fixes. A third fixed point corresponds to a community where $(x_A,x_{AB},x_B)=(\frac{1}{2}, x_{AB}^*, \frac{1}{2})$ where
\begin{equation}
\label{unstable}
x_{AB}^\ast = \left\{ \begin{array}{ll} \frac{1}{2(1+\mu)} & \mbox{if $b=0$ or $c = 0$} \\ \frac{1+\mu}{bc} \left[ \sqrt{1+\frac{bc}{(1+\mu)^2}} -1 \right] & \mbox{otherwise} \end{array} \right. \;,
\end{equation}
and which is stable when $\mu<0$. Unless $x_{AB}^*=1$, this community exhibits a mixture of $AA$, $AB$ and $BB$ agents. The $AA$ and $BB$ agents coexist in equal numbers, whilst $x_{AA}=x_{BB}=\frac{1}{2}(1-x_{AB}^\ast)$. Note that this mixed state comprising consistent and inconsistent agents is distinct from that found in models where agents that are too dissimilar (e.g., $AA$ and $BB$ agents) cannot interact (e.g., \cite{vaz04}). In those models, one finds a frozen state where agents do not change their behaviour over time. Here, by contrast, the agents perpetually change state since, for example, if an $AA$ agent meets a $BB$, one of them will be an $AB$ after the interaction.

A state with this inconsistent (or `undecided') character was also seen in the generalisation of the Naming Game studied in \cite{bar07}. This generalisation introduces to the basic Naming Game a probability $p$ of updating the agents' states after a successful interaction\footnote{In \cite{bar07} this parameter is called $\beta$, but we use a different symbol here to avoid a clash of notation}. The choice $p=1$ recovers the basic Naming Game dynamics. By constructing the set of transition probabilities, analogous to Table~\ref{rates}, one finds that the microscopic dynamics for general $p$ cannot be realised with a judicious choice of our parameters $b$ and $c$. However, the resulting deterministic equations of motion coincide with (\ref{xA})--(\ref{xB}) if one takes
\begin{eqnarray}
 \mu &=& \frac{b+c}{2} = \frac{3p - 1}{2} \\
 bc &=& p \;,
\end{eqnarray}
that is,
\begin{equation}
 c = \frac{1+b}{3b-1} \;.
\end{equation}
Thus within the allowed range of b, $-1<b<1$, the deterministic dynamics of the Naming Game generalisation described in [22] can be replicated only if $c$ is allowed to take on unphysical values ($c>1$ or $c<0$). Nevertheless, the collective dynamics of both generalisations (as described in [22] and the following sections) turn out to be quite similar. We return to this point in the discussion.

\section{Simplified dynamics}
\label{simple}

\begin{figure}
\includegraphics[width=0.33\linewidth]{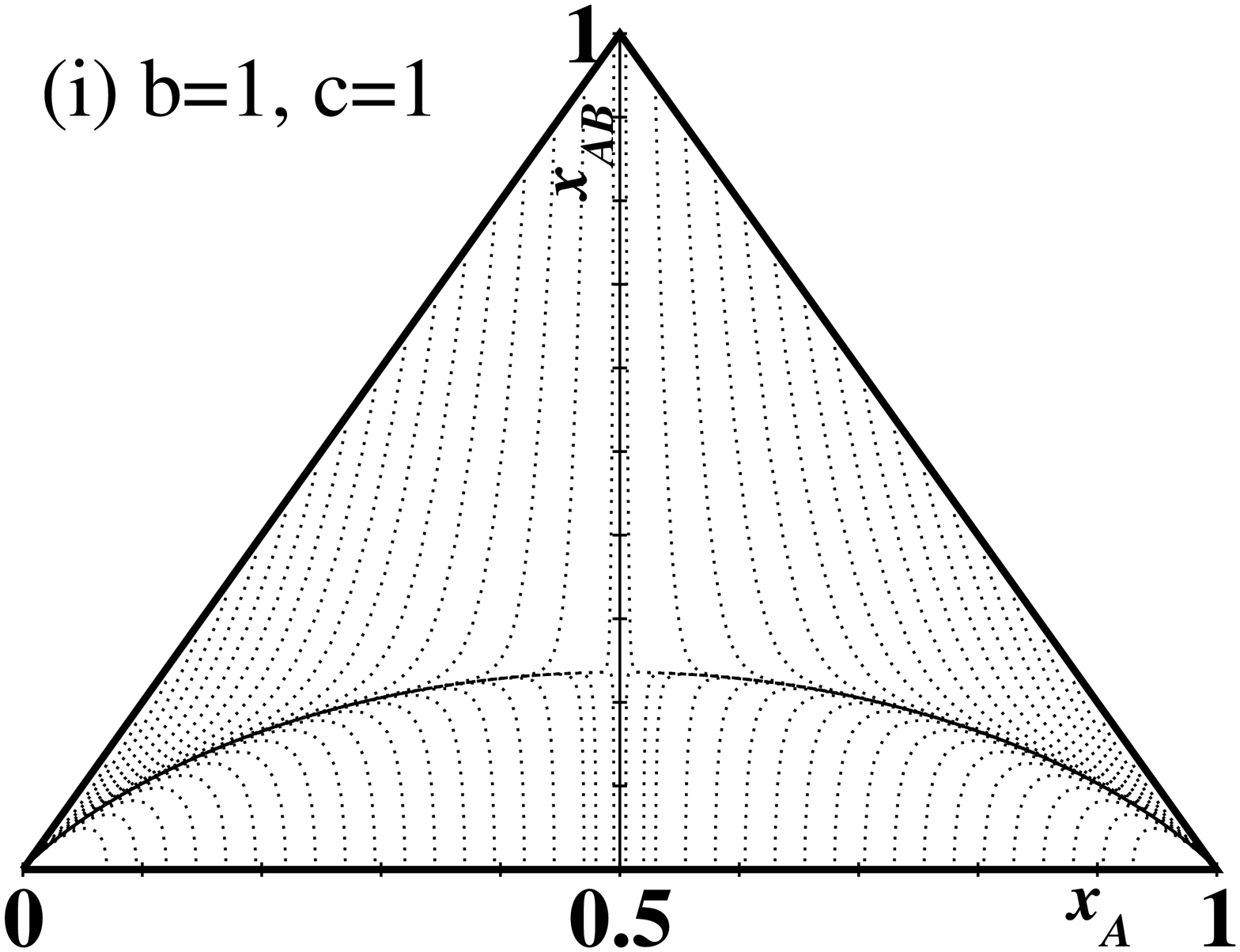}
\includegraphics[width=0.33\linewidth]{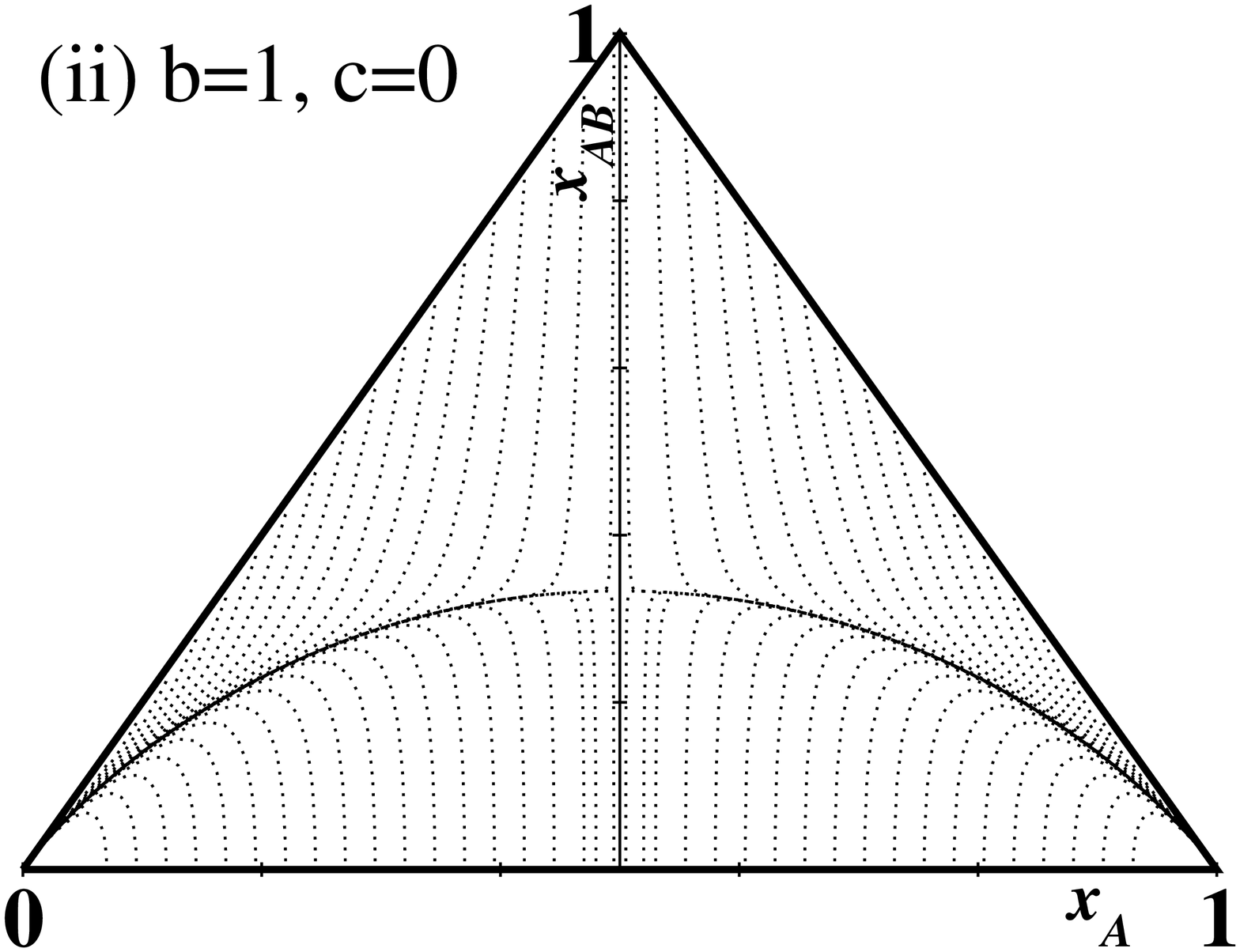}
\includegraphics[width=0.33\linewidth]{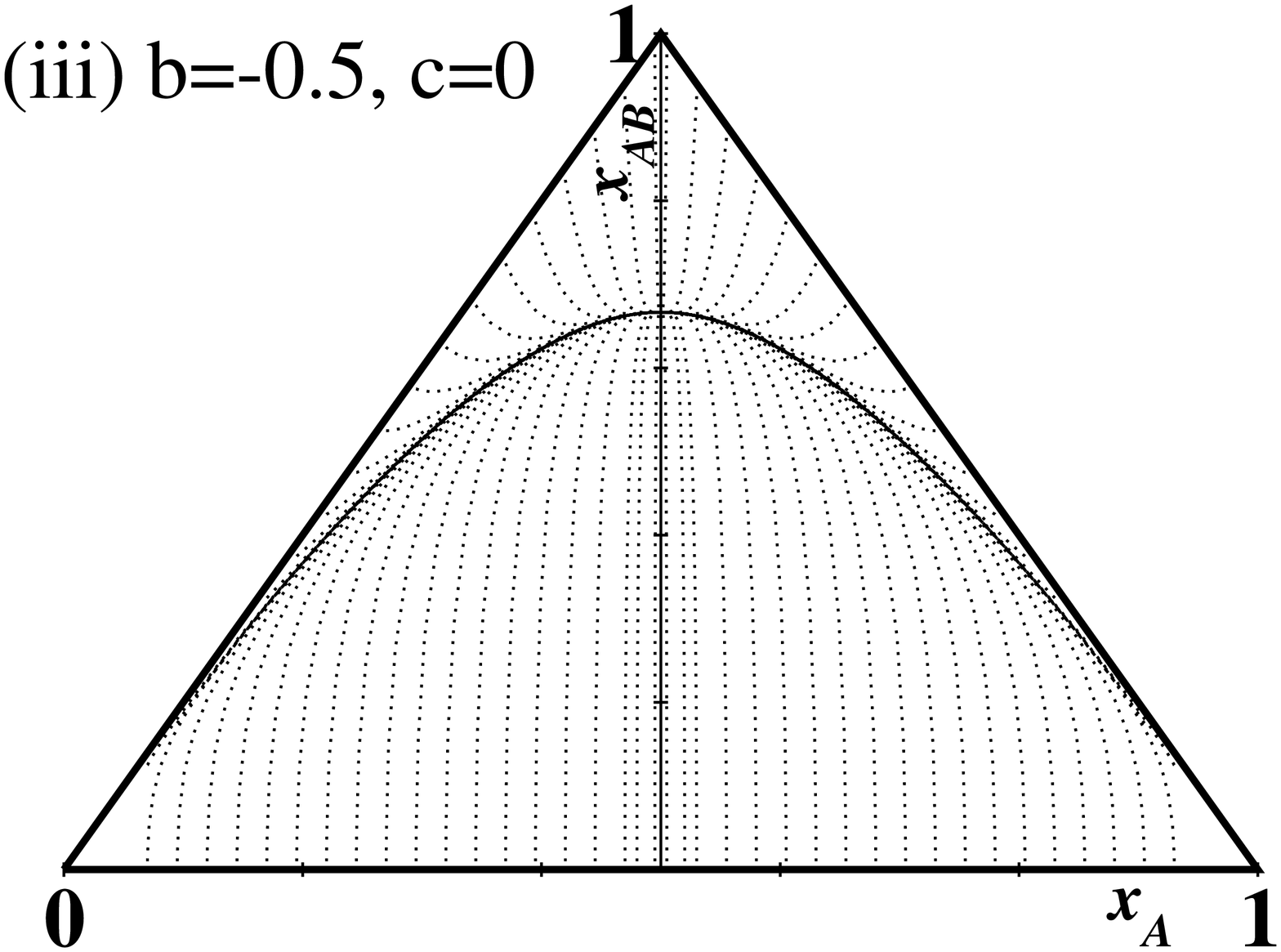}
\caption{\label{htrajs} Trajectories of the deterministic system (\ref{xA})--(\ref{xB}) obtained using a numerical Runge-Kutta-Fehlberg (RKF45) algorithm \cite{GSL}. In cases (i) and (ii), $\mu>0$ so the fixed points at $(x_A,x_{AB})=(0,0)$ and $(0,1)$ are attractors of the dynamics, whereas in case (iii) $\mu<0$ and all solutions approach the fixed point at $(\frac{1}{3},\frac{2}{3})$.}
\end{figure}

Our main aim in this work is to understand how the stochastic component of the dynamics affects the picture described in the previous section. Given the complexity of the deterministic equations (\ref{xA})--(\ref{xB}), it seems unlikely that their stochastic counterparts will be analytically tractable. Our strategy is to use insights from the deterministic dynamics to design a simplified dynamics that can be couched in terms of a single stochastic coordinate. Then, the resulting Fokker-Planck equation can be studied using more-or-less standard methods \cite{red01,ewe04}.

We begin by examining numerical solutions of the system of equations (\ref{xA})--(\ref{xB}). These are plotted in Fig.~\ref{htrajs} for three combinations of the parameters $b$ and $c$ in the $x_A$--$x_{AB}$ plane.  What is striking is that all solutions begin with a trajectory whereby the number of $AB$ agents changes whilst the number of $A$ and $B$ tokens remains roughly constant. Then, a common curve $x_{AB}=f(x_A)$ is followed towards the fixed point. A look at the time series (not shown), suggests that the time taken to reach this curve is typically less than 10\% of the time taken to reach the fixed point. Therefore, within the simplified dynamics, we will assume that $x_{AB}\approx f(x_A)$ for the entire trajectory. This will then yield an equation of motion for a single coordinate $x_A$. Unfortunately, we have not been able to find an analytic expression for the curve $f(x_A)$. However, it does not deviate too far from the parabola that passes through $(x_A,x_{AB})=(0,0),(\frac{1}{2},x_{AB}^\ast),(1,0)$, viz, $f(x_A) = 2\alpha x_A(1-x_A) = 2\alpha x_A x_B$, where we have introduced a second important parameter
\begin{equation}
\label{alpha}
 \alpha = 2 x_{AB}^\ast \;.
\end{equation}

This approximation can be interpreted in the following way. Suppose first of all that there is no correlation between the two tokens in each agent's store: given a randomly chosen agent, the first token is $A$ with probability $x_A$, and so is the second. The probability that an agent is inconsistent is then $2 x_A x_B$. Thus, the choice $\alpha=1$ is equivalent to a mean-field type approximation in which we assume that the two tokens in each agent's store are uncorrelated. Values of $\alpha$ different from $1$ imply correlations between the pair of tokens contained in a randomly-chosen agent's store. Specifically, the probability of an agent being in the inconsistent state is $x_{AB}-2x_Ax_B = 2(1-\alpha)$ more likely than chance. Alternatively, if one of the stored tokens is an $A$, the other is an $A$ with probability $1-\alpha x_B$; likewise, if one is a $B$, so is the other with probability $1-\alpha x_A$.

This latter interpretation provides a means to simulate the simplified dynamics directly. One maintains a collective token store with (notionally) $m_A = 2N x_A$ instances of $A$ and $m_B = 2N x_B$ instances of $B$. Then, one constructs a speaker in state $AA$ with probability $x_A(1-\alpha x_B)$, in state $BB$ with probability $x_B(1-\alpha x_A)$, or $AB$ otherwise. Likewise a listener. Then speaker and listener interact, and the token frequencies change probabilistically according to transition probabilities and the $\delta m$ values given in Table~\ref{rates}. When $\mu<0$, it turns out that $\alpha>1$, which in turn means that the probability of constructing a speaker in a consistent state becomes negative when the corresponding token frequency $x<1-\frac{1}{\alpha}$. In the event that this occurs, one can simply set the offending probability to zero.

It is not \textit{a priori} clear whether the choice $\alpha=2x_{AB}^\ast$ yields a dynamics that is more faithful to the full hybrid model dynamics than the mean-field approximation $\alpha=1$. This we will establish by comparing predictions for these two choices of $\alpha$ with data from simulations of the full stochastic dynamics (see Section~\ref{sim}).

\section{Stochastic equations of motion for the simplified dynamics}
\label{stoch}

The stochastic equations of motion are obtained by evaluating in addition to (\ref{det}) the mean of the square change in token frequencies in the course of an interaction. This is evaluated by computing a sum over all possible transitions with the help of Table~\ref{rates}, analogous to the procedure used to derive (\ref{det}). Since we have reduced to a model whose state is given by a single parameter $x_{A}$, we only need to calculate
\begin{equation}
\label{sto}
 \langle (\delta x_A)^2 \rangle = \frac{\langle (\delta m_A)^2 \rangle}{4 N^2} = \frac{1}{4N^2} \left[ 2 x_A x_B + \mu x_{AB} + \left(1 + \frac{3}{2}b\right) c x_{AB}^2 \right] \;.
\end{equation}

Using (\ref{det}), (\ref{sto}) and approximating $x_{AB}$ as $2\alpha x_A(1-x_A)$, we obtain a Fokker-Planck equation \cite{ris89} for the simplified dynamics:
\begin{eqnarray}
\label{fpe1}
\frac{\partial}{\partial t} P(x,t) &=&  \frac{\mu \alpha}{N} \frac{\partial}{\partial x}  x(1-x)(1-2x) P  + \nonumber\\ &&\qquad \frac{1}{4N^2} \frac{\partial^2}{\partial x^2} \left[ (1+\mu\alpha) x(1-x) + \alpha^2(2+3b)c x^2(1-x)^2 \right] P 
\end{eqnarray}
where we have replaced $x_A$ with $x$ to lighten the notation. Introducing a rescaled time variable
\begin{equation}
\tau = \frac{\mu\alpha}{N} t
\end{equation}
and the two parameters
\begin{eqnarray}
\beta &=& \frac{4 \mu \alpha N}{1 + \mu\alpha} \\
\sigma &=& \frac{\alpha^2}{4} \frac{(2+3b)c}{1+\mu\alpha}
\label{sigma}
\end{eqnarray}
we may rewrite the Fokker-Planck equation as
\begin{equation}
\label{fpe}
\frac{\partial}{\partial \tau} P(x,\tau) = \frac{\partial}{\partial x}  x(1-x)(1-2x) P  + \frac{1}{\beta} \frac{\partial^2}{\partial x^2}  x(1-x) \left[ 1 + 4 \sigma x(1-x) \right] P  \;,
\end{equation}
where the factors of $4$ are for future convenience. The parameter $\beta$ plays the role of inverse temperature, controlling the magnitude of the stochastic term. It is proportional to the community size $N$, indicating that a large community exhibits, if $|\mu \alpha| \gg 1/N$, a low-temperature behaviour in which stochastic effects are expected to act as a perturbation about the deterministic dynamics.

It is helpful briefly to transform the Fokker-Planck equation from the form
\begin{equation}
\label{fpex}
\frac{\partial}{\partial \tau} P(x,\tau) = - \frac{\partial}{\partial x} a(x) P + \frac{1}{\beta} \frac{\partial^2}{\partial x^2}  b(x) P
\end{equation}
via a change of variable $x \to y(x)$ to the form
\begin{equation}
\label{fpey}
\frac{\partial}{\partial \tau} P(y,\tau) = \frac{\partial^2}{\partial y^2}  \tilde{b}(y) P \;.
\end{equation}
Making this change of variable in the derivatives of (\ref{fpex}) reveals that $y(x)$ is required to satisfy
\begin{equation}
\frac{{\rm d} y}{{\rm d} x} a(x) + \frac{1}{\beta} \frac{{\rm d}^2 y}{{\rm d} x^2} b(x) = 0
\end{equation}
so that the first-order term vanishes. This is achieved by setting
\begin{equation}
\label{by}
\tilde{b}(y) = \frac{1}{\beta} \left( \frac{{\rm d} y}{{\rm d} x} \right)^2 b(x) \;.
\end{equation}
Defining the `potential' $V(x)$ through
\begin{equation} 
\label{aVb}
a(x) = - \frac{{\rm d} V}{{\rm d} x} b(x) \;,
\end{equation}
which here means
\begin{equation}
\label{pot}
V(x) = \frac{1}{4 \sigma} \ln\left[ 1 + 4 \sigma x(1-x) \right]
\end{equation}
(which approaches $V(x) = x(1-x)$ as $\sigma\to0$), it then follows that
\begin{equation}
\frac{{\rm d}y}{{\rm d}x} = A {\rm e}^{\beta V(x)} \;.
\end{equation}
The constants of integration we fix by mapping the interval $x \in [0,1]$ to $y \in [-1,1]$ in such a way that $y(1-x) = -y(x)$. Then,
\begin{equation}
\label{yx}
y(x) = \frac{2 \int_{\frac{1}{2}}^x {\rm d}u\, {\rm e}^{\beta V(u)}}{\int_0^1 {\rm d}u\, {\rm e}^{\beta V(u)}} \;.
\end{equation}

The value of this transformation is that \emph{first passage properties} \cite{red01} can be found relatively straightforwardly from the backward equation corresponding to (\ref{fpey}). Specifically, the probability $Q(x)$ that the boundary at $x=1$ is the first boundary to be encountered by the dynamics is give in terms of the transformed coordinate through the solution of
\begin{equation}
\tilde{b}(y) \frac{{\rm d}^2}{{\rm d} y^2} \tilde{Q}(y) = 0
\end{equation}
subject to the boundary conditions $\tilde{Q}(-1)=0$ and $\tilde{Q}(1)=1$ \cite{red01,ewe04}. The solution is
\begin{equation}
\label{Qy}
\tilde{Q}(y) = \frac{1+y}{2} \quad\Longrightarrow\quad Q(x) = \frac{1+y(x)}{2} \;.
\end{equation}
Meanwhile, the mean time to reach the boundary $x=1$ starting from a position $x$ \emph{given} that this is the first boundary to be encountered is given by the solution of
\begin{equation}
\tilde{b}(y) \frac{{\rm d}^2}{{\rm d} y^2} \tilde{Q}(y) \tilde{T}(y) = - \tilde{Q}(y) \;,
\end{equation}
subject to the boundary conditions that $\tilde{T}(1) = 0$ and $\tilde{T}(-1) < \infty$ \cite{red01,ewe04}. We find
\begin{equation}
\tilde{T}(y) = \tilde{I}(1) - \tilde{I}(y) \quad\mbox{where}\quad
\tilde{I}(y) = \frac{1}{1+y} \int_{-1}^y {\rm d} u \, \frac{(y-u)(1+u)}{\tilde{b}(u)} \;,
\label{Iy}
\end{equation}
as long as $\tilde{I}(\epsilon) \sim \epsilon$ as $\epsilon \to 0$ (a condition satisfied here). In terms of the original variable $x$, this solution reads
\begin{equation}
\label{Tx}
T(x) = I(1) - I(x)
\end{equation}
where
\begin{equation}
\label{Ix}
I(x) = \frac{\beta}{2} \frac{\int_0^1 {\rm d}u\, {\rm e}^{\beta V(u)}}{1+y(x)} \int_{0}^x {\rm d} u \, {\rm e}^{-\beta V(u)} \frac{\left[y(x)-y(u)\right] \left[1+y(u)\right]}{u(1-u)[1+4\sigma u(1-u)]} \;.
\end{equation}
Analysis of these results will be performed separately within the three distinct regimes exhibited by model.

\section{Consensus properties of the simplified dynamics in large communities}
\label{large}

In this section we analyse the large community-size (large-$N$) properties of the probability and mean time to reach consensus on the variant $A$ within the simplified dynamics, given an initial frequency $x$ of $A$ tokens. At any fixed $\mu$, one will access for sufficiently large $N$ a large-$\beta$ regime that is dominated by the deterministic dynamics: recall that the parameter $\beta$, which is roughly proportional to $\mu N$, plays the role of inverse temperature. Nevertheless, stochastic effects make their presence felt, even in this low-temperature regime, as we will see below. On the other hand, if $\mu$ scales with $N$ as $\mu \sim 1/N$, then both deterministic and stochastic effects are of a similar magnitude. Here we can gain some insights into the model's dynamics by perturbing around the purely stochastic case, $\mu=0$, where relaxation to consensus is dominated by a diffusive mode. Comparison of these analytical results with simulations will follow in the next section.

\subsection{Low-temperature behaviour of the rapid consensus phase, $\mu>0$}

When $\mu$ is positive, so is $\beta$, and integrals of the type appearing in (\ref{yx}) are dominated by the maximum in $V(x)$ at $x=\frac{1}{2}$. To evaluate $y(x)$, we note that by writing $x = \frac{1}{2}+\frac{u}{\sqrt{\beta}}$, 
\begin{equation}
\beta V(x) = \frac{\beta}{4\sigma} \ln \left( 1 + \sigma \right) - \frac{1}{1+\sigma} u^2 + O\left(\frac{1}{\sqrt{\beta}}\right) \;.
\end{equation}
In particular this implies that the integral
\begin{equation}
 \label{prefactor}
 \int_0^1 {\rm d} u {\rm e}^{\beta V(u)} \sim \sqrt{\frac{\pi (1+\sigma)}{\beta}} {\rm e}^{\beta V(\frac{1}{2})}
\end{equation}
and, in the central region of width $\sim 1/\sqrt{\beta}$, that
\begin{equation}
\label{yerf}
 y(x) \sim {\rm erf}\left[ \sqrt{\frac{\beta}{1+\sigma}} \left( x - \frac{1}{2} \right) \right]
\end{equation}
with corrections of order $1/\sqrt{\beta}$. Outside this central region we may instead write
\begin{eqnarray}
 y(x) &=& 1 - \frac{ \int_{x}^1 {\rm d}u\, {\rm e}^{\beta[ V(u) - V(1/2) ]} }{ \int_{1/2}^1 {\rm d}u\, {\rm e}^{\beta[ V(u) - V(1/2) ]} } \\
 &\sim& 1 - \frac{2\sqrt{\beta}}{\sqrt{\pi(1+\sigma)}} \int_x^1 {\rm d} u\, {\rm e}^{\beta [V(u) - V(1/2)]}\\
 &\sim& 1 + \frac{2 {\rm e}^{-\beta V(1/2)}}{\sqrt{\pi \beta(1+\sigma)}} \left[ \frac{{\rm e}^{\beta V(x)}}{V'(x)} +  1 \right] + O\left( \frac{1}{\beta} \right) \qquad \mbox{when $x \gg \frac{1}{2}$}
 \label{yasymp}
\end{eqnarray}
where, to arrive at the last line we inserted $1=\frac{V'}{V'}$ into the integrand and integrated by parts, and further used that $V(1)=0$ and $V'(1)=-1$. A similar result is obtained for $x \ll \frac{1}{2}$ by invoking the antisymmetry property
$y(1-x)=-y(x)$:
\begin{equation}
y(x) \sim -1 + \frac{2 {\rm e}^{-\beta V(1/2)}}{\sqrt{\pi \beta(1+\sigma)}} \left[ \frac{{\rm e}^{\beta V(x)}}{V'(x)} -  1 \right]  + O\left( \frac{1}{\beta} \right) \qquad \mbox{when $x \ll \frac{1}{2}$.}
\label{yasympleft}
\end{equation}
An expression for the consensus probability then follows from (\ref{Qy}).

We now turn to mean consensus time, given analytically for the simplified dynamics by the integral (\ref{Ix}). Evaluation of this integral is a somewhat more involved enterprise than it was for the consensus probability. To avoid an excessively tedious presentation, we will omit some steps of routine algebra and focus on the main ideas behind the derivation.

We anticipate that the mean consensus time from a given initial condition $x$ will be an increasing function of the number of agents, $N$, and hence $\beta$. Therefore, we will systematically drop any terms in (\ref{Ix}) that vanish in the limit $\beta\to\infty$. For example, consider the regime $0 < x < \frac{1}{2}$, where by the second inequality we mean $\frac{1}{2} - x \gg \frac{1}{\sqrt{\beta}}$.  At the bottom end of the integral, we have that $[1+y(u)]/[1+y(x)]$ vanishes exponentially fast with $\beta$. Hence, in (\ref{Ix}), the combination behaves as
\begin{equation}
 \frac{y(x)-y(u)}{1+y(x)} = \frac{[1+y(x)] - [1+y(u)]}{1+y(x)} \sim 1 + \epsilon_{\beta} \;,
\end{equation}
in which $\epsilon_\beta$ is an exponentially small correction and will hence be neglected. Using (\ref{prefactor}) and (\ref{yasympleft}), one can further show that
\begin{equation}
\frac{\beta}{2} \int_0^1 {\rm d}v\, {\rm e}^{\beta V(v)} \left[ 1 + y(u) \right] \sim \frac{{\rm e}^{\beta V(u)}}{V'(u)} - 1 \;.
\end{equation}
Substituting into (\ref{Ix}), and using (\ref{aVb}), we find that
\begin{equation}
\label{Ixlo}
 I(x) \sim \int_0^x \frac{{\rm d} u}{u(1-u)(1-2u)} \left[ 1 - V'(u) {\rm e}^{-\beta V(u)} \right] \;,
\end{equation}
for $x<\frac{1}{2}$. In principle we should add to this contributions to (\ref{Ix}) from the top end, $u=x$. However, by using (\ref{yasympleft}), and taking into account that $1+y(u)$ and $1+y(x)$ are of comparable magnitude in this regime, one ultimately finds that these contributions vanish in the limit $\beta\to\infty$. 

This expression has an interesting interpretation. Since the mean time to reach the right boundary $x=1$ (conditioned on that being the first boundary that is encountered) satisfies $T(x_1)-T(x_2) = I(x_2)-I(x_1)$ we have, if $x_1$ and $x_2$ are both sufficiently far away from the left boundary that the term ${\rm e}^{-\beta V(u)}$ can be neglected,
\begin{equation}
\label{Tdiff}
\fl T(x_1) - T(x_2) = I(x_2)-I(x_1) \sim \int_{x_1}^{x_2} \frac{{\rm d} u}{u(1-u)(1-2u)} = \left[ \ln \frac{u(1-u)}{(1-2u)^2} \right]_{x_1}^{x_2}\;.
\end{equation}
The right-hand side of this expression can be recognised as the integral $\int_{x_2}^{x_1} \frac{{\rm d} u}{a(u)}$, where $a(u)$ is the deterministic `force' term that appears in the Fokker-Planck equation (\ref{fpe}). In a deterministic interpretation, this is the time derivative of $x$, and so the integral gives the time taken for the deterministic dynamics to reach $x_1$ from the point $x_2>x_1$ (assuming $x_2<\frac{1}{2}$). We see then that, when the stochastic dynamics are conditioned on being absorbed at the boundary $x=1$, the additional time needed to traverse the interval $[x_1,x_2]$ is on average equal to that required by the deterministic dynamics, even though this pushes the coordinate $x$ in the opposite direction to the stochastic fluctuation!

In the deterministic dynamics, one cannot obtain the consensus time through the limit $x_2 \to 0$, as the integral diverges. Typically, given the underlying discrete nature of the process, one imposes the cut-off $x_2 = 1/N$. Within the stochastic dynamics this cut-off is handled automatically by the term in the square brackets of (\ref{Ixlo}).  Integrating by parts,
\begin{eqnarray}
 I(x) &\sim& \left[ \ln \frac{u(1-u)}{(1-2u)^2} \left(1 - V' {\rm e}^{-\beta V}\right) \right]_0^x + {} \nonumber\\ && \qquad \int {\rm d}u\, \ln \frac{u(1-u)}{(1-2u)^2} \left( V'' - \beta V' \right) {\rm e}^{-\beta V} \\
&\sim& \ln \frac{x(1-x)}{(1-2x)^2} - \int_0^\infty {\rm d}v\, \ln \left( \frac{v}{\beta} \right) {\rm e}^{-v} \\
&\sim& \ln \frac{x(1-x)}{(1-2x)^2} + \ln \beta + \gamma
\label{Ixleft}
\end{eqnarray}
where $\gamma=- \int_0^\infty {\rm d}x \ln x {\rm e}^{-x} = 0.5722\ldots$ is the Euler-Mascheroni constant (see \cite{gra00} \S 4.331). In the second line we made the change of variable $v=\beta u$, expanded $V$ and its derivatives around $x=0$ and discarded terms of order $1/\beta$ and lower.

In the Appendix, we show that the relation
\begin{equation}
\label{Isym}
\lim_{\beta\to\infty} \left[ I(x) + I(1-x) - I(1) \right] = 0
\end{equation}
holds in the range $|x - \frac{1}{2}| \gg \frac{1}{\sqrt{\beta}}$. It then automatically follows that the consensus time from an initial coordinate $x \gg \frac{1}{2}$, has as its leading terms $T(x)=I(1)-I(x) = I(1-x)$, i.e., the expression (\ref{Ixleft}) with $x \to 1-x$. Then to find the mean consensus time for a minority variant, $x\ll\frac{1}{2}$, all that is needed is an expression for $I(1)$.

In this one remaining integral, the contribution from the range $u \in [0,x]$ when $\frac{1}{2} - x \ll \frac{1}{\sqrt{\beta}}$ is given by (\ref{Ixleft}). Likewise, by symmetry of the integrand, the range $u \in [1-x,1]$ contributes the same amount. To estimate the contribution from the central part of the integral, we make the change of variable $u=\frac{1}{2} + \sqrt{(1+\sigma)/\beta} v$. The central part of the integral then approaches
\begin{equation}
 \sqrt{\pi} \int_{-v_0}^{v_0} {\rm d} v {\rm e}^{v^2} \left[ 1 - {\rm erf}^2 (v) \right]
\end{equation}
as $\beta\to\infty$ with $v_0$ fixed (so that the range over which the original integral was performed becomes ever narrower).  In this central region we are justified in using the error function (\ref{yerf}) as an approximation to $y(x)$.  At large arguments $\sqrt{\pi} {\rm e}^{v^2} [1 - {\rm erf}(v)] \sim 1/v$, and so we anticipate that for large $v_0$ this integral grows as $4 \ln v_0 + \Gamma$, where $\Gamma$ is a constant we estimate by numerical integration (up to $v_0=200$) as $\Gamma\approx2.541$. Substituting $x = \frac{1}{2} + \sqrt{(1+\sigma)/\beta} v_0$ into (\ref{Ixleft}), and adding this central contribution we find that the terms involving $v_0$ cancel leaving
\begin{equation}
 I(1) \sim 4 \ln \beta + 2 \gamma + \Gamma - 4 \ln 2 - 2 \ln(1+\sigma) 
\end{equation}
plus corrections that vanish with $\beta$.

We may now summarise the forms of the mean time $T(x)$ to reach consensus on $A$, given that initially a fraction $x$ of all tokens were of $A$ as
\begin{equation}
\label{Ts}
 T(x) \sim \left\{ \begin{array}{ll}
               4\ln \beta + 2\gamma - \delta - 2 \ln(1+\sigma) & x \to 0 \\
	       3\ln \beta + \gamma - \delta - 2 \ln(1+\sigma) - \ln \frac{x(1-x)}{(1-2x)^2} & 0 \ll x \ll \frac{1}{2} \\
	       \ln \beta + \gamma + \ln \frac{x(1-x)}{(2x-1)^2} & \frac{1}{2} \ll x \ll 1 \\
	       0 & x \to 1  
                \end{array}
	\right.
\end{equation}
where $\delta = 4 \ln 2 - \Gamma \approx 0.2316$ and $\gamma \approx 0.5722$.

Recall that we have rescaled time so that one unit of time corresponds to $N/\mu\alpha$ interactions between pairs of agents. Recall also that $\beta$ is proportional to the number of agents $N$. Therefore, to reach consensus on $A$, conditioned on this being the final outcome, each agent must interact of order $\ln N$ times (ignoring prefactors), no matter the initial frequency of $A$ tokens. Na\"{\i}vely, one might expect that the time to reach consensus on $A$ should increase dramatically from an initial condition in which it is in the minority. If the frequency $x$ is interpreted as a particle coordinate, it has to overcome a `potential barrier' at $x=\frac{1}{2}$. This event we expect to be exponentially rare in the low-temperature (large-$\beta$) limit, which it is, and hence take an exponentially long time to occur on average. This would be true if it were not for the fact that absorption takes place at the boundaries $x=0,1$. To reach the boundary $x=1$ from $x<\frac{1}{2}$, the particle needs to avoid the boundary $x=0$. It seems likely that this is achieved through an early-time fluctuation over the barrier to the region $x>\frac{1}{2}$. This would then provide a mechanism for all consensus times being of order $\ln N$.

\subsection{Low-temperature behaviour of the delayed-consensus phase: $\mu<0$}

We turn now to the case of $\mu<0$ where, in the deterministic limit, an inconsistent state (no consensus) is always reached. The effect of noise is to allow the consistent absorbing state to be reached. Hence then, the inconsistent state is a metastable state, and the two consistent states are the true steady states of the dynamics.

When $\mu<0$, the earlier rescaling of time to arrive at (\ref{fpe}) would involve our taking time to decrease towards $-\infty$ as the system involves. Since this is somewhat confusing, we make instead the change of variable
\begin{equation}
 \tau = \frac{|\mu|\alpha}{N} t
\end{equation}
which leads to the Fokker-Planck equation
\begin{equation}
 \frac{\partial}{\partial \tau} P(x,\tau) = - \frac{\partial}{\partial x}  x(1-x)(1-2x) P  + \frac{1}{|\beta|} \frac{\partial^2}{\partial x^2}  x(1-x) \left[ 1 + 4 \sigma x(1-x) \right] P  \;,
\end{equation}
where we note that if $\mu$ is negative, then so is $\beta$. We see that the effect of changing the sign of $\mu$ is to flip the sign of the deterministic term. That is, a particle now has to diffuse out of a potential well centred on $x=\frac{1}{2}$ to reach a boundary point $x=0,1$. Here, the physical intuition that the time to reach the boundary should increase exponentially with $|\beta|$ and the well depth is valid. Furthermore, one anticipates that the strong attraction towards the potential minimum that occurs when $|\beta|$ is large leads to the initial condition being forgotten, and either boundary ultimately being reached with equal probability. The exception would be for a particle starting close to a boundary, and which experiences an early-time fluctuation that leads to almost immediate absorption.

We now confirm these expectations explicitly. Recall that the transformed coordinate $y(x)$ is given by the integral (\ref{yx}) which can be written as
\begin{equation}
y(x) = \frac{\int_{\frac{1}{2}}^{x} {\rm d} u {\rm e}^{-|\beta| V(u)}}{\int_{\frac{1}{2}}^1 {\rm d}u {\rm e}^{-|\beta| V(u)} }
\end{equation}
when $\beta$ is negative. Here the function $V(x)$ is still as given by the positive function (\ref{pot}); we write the minus signs that appear explicitly. Both integrals are dominated by their upper end-point, i.e.,
\begin{equation}
\label{negint}
\int_{\frac{1}{2}}^{x} {\rm d} u {\rm e}^{\beta V(u)} \sim {\rm e}^{\beta V(x)} \left( \frac{1}{\beta V'(x)} + O(1/|\beta|^2) \right) \;.
\end{equation}
Hence, when $|x-\frac{1}{2}| \gg \frac{1}{\sqrt{|\beta|}}$, we have asymptotically that
\begin{equation}
\label{ynegasymp}
y(x) \sim  \frac{1+4\sigma x(1-x)}{2x-1} {\rm e}^{-\frac{|\beta|}{4\sigma} \ln[1+4\sigma x(1-x)]} \;.
\end{equation}
Notice that $y(x)$ differs significantly from zero only in a region of size $\frac{1}{|\beta|}$ near the boundary points $x=0$ and $x=1$. Therefore, the probability of consensus on $A$ from initial frequency $x$ is, via the expression (\ref{Qy}), $Q(x) = \frac{1}{2}[1+y(x)]$, close to one half for all $x$ except in these boundary regions, as previously claimed.

The integral $I(x)$, (\ref{Ix}), that appears in the formula $T(x) = I(1)-I(x)$ for consensus time, (\ref{Tx}), can be written as
\begin{equation}
 I(x) \sim \frac{1}{1+y(x)} \int_0^x {\rm d}u\; {\rm e}^{|\beta| V(u)} \frac{[y(x)-y(u)][1+y(u)]}{u(1-u)[1+4\sigma u(1-u)]} \;,
\end{equation}
where we have used (\ref{negint}) to evaluate the first integral in (\ref{Ix}) to leading order in $|\beta|$. Note that the units of time here are those introduced earlier in this Section.

If $x-\frac{1}{2} \gg 1/\sqrt{|\beta|}$, the integral is dominated by the midpoint $u=\frac{1}{2}$ where the function $V(x)$ has a maximum. In this region, $y(u)$ is roughly constant and close to zero. Hence, to leading order in $|\beta|$,
\begin{eqnarray}
 I(x) &\sim& - \frac{4}{1+\sigma} \frac{y(x)}{1+y(x)} {\rm e}^{|\beta| V(\frac{1}{2})} \int_0^x {\rm d}u\; {\rm e}^{-\frac{1}{2} |\beta V''(\frac{1}{2})| (u-\frac{1}{2})^2} \\
&\sim& - \sqrt{\frac{16\pi}{|\beta|(1+\sigma)}} \frac{y(x)}{1+y(x)} {\rm e}^{\frac{|\beta|}{4\sigma} \ln(1+\sigma)} \;.
\end{eqnarray}
Contributions from either end-point are subdominant, and we see that because $y(x)$ decays rapidly to zero near $x=1$, the dominant contribution to the difference $T(x)=I(1)-I(x)$ is from $I(1)$ when the distance from the right boundary is much larger than $\frac{1}{|\beta|}$. When $x<\frac{1}{2}$, the mid-point does not contribute at all, and one has only the subleading end-point contributions to the integral, and hence one finds that for sufficiently large $|\beta|$, the consensus time is roughly constant over the intermediate range of $x$. That is,
\begin{equation}
\label{Txneg}
 T(x) \sim I(1) \sim \sqrt{\frac{4\pi}{|\beta| (1+\sigma)}} {\rm e}^{\frac{|\beta|}{4\sigma} \ln(1+\sigma)} \;,
\end{equation}
where we used the fact that $y(x)\to 1$ as $x\to1$.  Note that in the limit $\sigma\to0$, we have
\begin{equation}
 T(x) \sim \sqrt{\frac{4\pi}{|\beta|}} {\rm e}^{\frac{|\beta|}{4}} \;.
\end{equation}
 This analysis thus confirms our expectation that the coordinate $x$ is typically pinned to a value of approximately $\frac{1}{2}$ for a time that increases exponentially with the inverse temperature $|\beta|$.

\subsection{The crossover regime: $|\mu| \sim 1/N$}

As previously discussed, one can define a crossover regime between the rapid- and delayed-consensus phases in a large community of size $N$ if the magnitude of $\mu \sim 1/N$. Then $\beta$ is of order unity and both the deterministic and stochastic terms in the Fokker-Planck equation (\ref{fpe}) are of similar magnitude.

To study this crossover, we take $\mu=\nu/N$, which in turn implies that both terms in the original Fokker-Planck equation (\ref{fpe1}) for the simplified dynamics are of order $1/N^2$. Thus here we should rescale time through
\begin{equation}
 \tau = \frac{t}{N^2} \;,
\end{equation}
so that then in the limit $N\to\infty$ the Fokker-Planck equation reads
\begin{equation}
\frac{\partial}{\partial \tau} P(x,\tau) = \nu\alpha \frac{\partial}{\partial x}  x(1-x)(1-2x) P  + \frac{1}{4} \frac{\partial^2}{\partial x^2}  x(1-x) \left[ 1 + 4 \sigma x(1-x) \right] P  \;,
\end{equation}
where, in this limit,
\begin{equation}
\sigma \to \frac{\alpha^2 (2+3b)c}{4} \;.
\end{equation}

There are two distinct cases that arise in the infinite-community limit, $N\to\infty$.

\medskip
\noindent\underline{Case I}: If both $b$ and $c$ vanish at least as fast as $1/N$, so that
\begin{equation}
\label{nulim}
 \nu = \lim_{N\to\infty} \frac{N(b+c)}{2}
\end{equation}
is not infinite, the parameter $\alpha = 2x_{AB}^\ast$, where $x_{AB}^\ast$ is given by (\ref{unstable}), tends to $1$ and $\sigma \to 0$. Then the Fokker-Planck equation simplifies to
\begin{equation}
 \frac{\partial}{\partial \tau} P(x,\tau) = \nu \frac{\partial}{\partial x}  x(1-x)(1-2x) P  + \frac{1}{4} \frac{\partial^2}{\partial x^2}  x(1-x) P  \;.
\end{equation}
Repeating the analysis following Eq.~(\ref{fpe}) leads to the following modified expressions for the transformed coordinate
\begin{equation}
 y(x) = \frac{2 \int_{\frac{1}{2}}^x {\rm e}^{4 \nu u(1-u)} }{ \int_0^1 {\rm d}u {\rm e}^{4 \nu u(1-u)}} \;,
\end{equation}
that appears in the consensus probability $Q(x)=\frac{1}{2}[1+y(x)]$ and the integral
\begin{equation}
 I(x) = 2 \frac{\int_0^1 {\rm e}^{4 \nu u(1-u)}}{1+y(x)} \int_0^x {\rm d}u {\rm e}^{-4 \nu u(1-u)} \frac{[y(x)-y(u)][1+y(u)]}{u(1-u)}
\end{equation}
that appears in the consensus time $T(x) = I(1)-I(u)$.

To get a feel for the crossover in this case, we expand about $\nu=0$ (the Voter Model) to first order in $\nu$ (although higher-order corrections could also be computed). We find
\begin{eqnarray}
\label{Qd}
 Q(x) &=& x + \frac{2}{3} x (1-x) (2x-1) \nu + O(\nu^2) \\
 T(x) &=& -4 \frac{(1-x) \ln (1-x)}{x} + {} \nonumber\\
 && \qquad \frac{1-x}{x} \left( \frac{8}{9} x (2x-7) - \frac{16}{3} (1-x) \ln(1-x)  \right) \nu + O(\nu^2) \;.
\label{Td}
\end{eqnarray}
As expected, consensus probability increases slightly for the majority variant when $\nu$ is small and positive. Intriguingly, to first order in $\nu$, consensus on the $A$ variant from any initial fraction of $A$ tokens is reduced relative to the purely diffusive case, $\nu=0$, when $\nu$ is positive. Again one might expect an increase in consensus times for minority variants. This we ascribe once again to the conditioning on trajectories that lead to consensus on $A$ when calculating $T(x)$. 

\medskip
\noindent\underline{Case II:} The other possibility is that $b$ and $c$ approach nonzero constants that have the same magnitude but opposite sign, i.e.,
\begin{equation}
b \sim - c^* + \frac{b'}{N^\delta} \quad\mbox{and}\quad c \sim c^\ast + \frac{c'}{N^\epsilon} \;,
\end{equation}
where $\delta, \epsilon \ge 1$. In this case, both $\alpha$ and $\sigma$ have nontrivial limiting values
\begin{equation}
 \alpha \to \frac{2}{(c^*)^2} \left( 1 - \sqrt{1 - (c^*)^2} \right) \quad\mbox{and}\quad \sigma \to \frac{\alpha^2(2-3c^*)c^*}{4} \;.
\end{equation}
The integrals for $y$ and $I$ are not as simple now:
\begin{eqnarray}
 y(x) &=& \frac{2 \int_{\frac{1}{2}}^{x} {\rm d}u {\rm e}^{\nu V(u)}}{\int_0^1 {\rm d}u {\rm e}^{\nu V(u)}} \\
 I(x) &=& \frac{2 \int_{0}^{1} {\rm d}u {\rm e}^{\nu V(u)}}{1 + y(x)} \int_0^x {\rm e}^{-\nu V(u)} \frac{[y(x)-y(u)][1+y(u)]}{u(1-u)[1+4\sigma u(1-u)]}
\end{eqnarray}
where here the potential is
\begin{equation}
 V=\frac{\alpha}{\sigma} \ln [ 1 + 4\sigma x(1-x) ] \;.
\end{equation}
In principle an expansion to first order in $\nu$ is possible; although in practice the resulting expressions are rather cumbersome and one does not gain any new insights.

\section{Comparison with Monte Carlo simulation}
\label{sim}

In deriving the analytical results of the previous section we have made two approximations: (i) that the community size $N$ is large, and thus that we are justified in keeping only the leading terms in an asymptotic expansion; and (ii) that the dynamics is well described by the stochastic evolution of a single coordinate, obtained through the simplification described in Section~\ref{simple}.  In this section, we examine the effect of these approximations independently by Monte Carlo simulation of (i) the simplified dynamics, using the algorithm outlined in Section~\ref{simple}, which will reveal any large finite-size contributions and (ii) the full dynamics, i.e., implementing the stochastic update rules presented in Table~\ref{rates}, which will show to what extent the simplified dynamics acts as a proxy for the full dynamics.  Our main findings are that consensus probabilities appear to be well predicted by the simplified dynamics for both models, but do not provide a complete quantitative account of the mean consensus time for the full dynamics outside the crossover regime.

\subsection{Finite-size behaviour of the simplified dynamics}

\begin{figure}
 \includegraphics[width=0.495\linewidth]{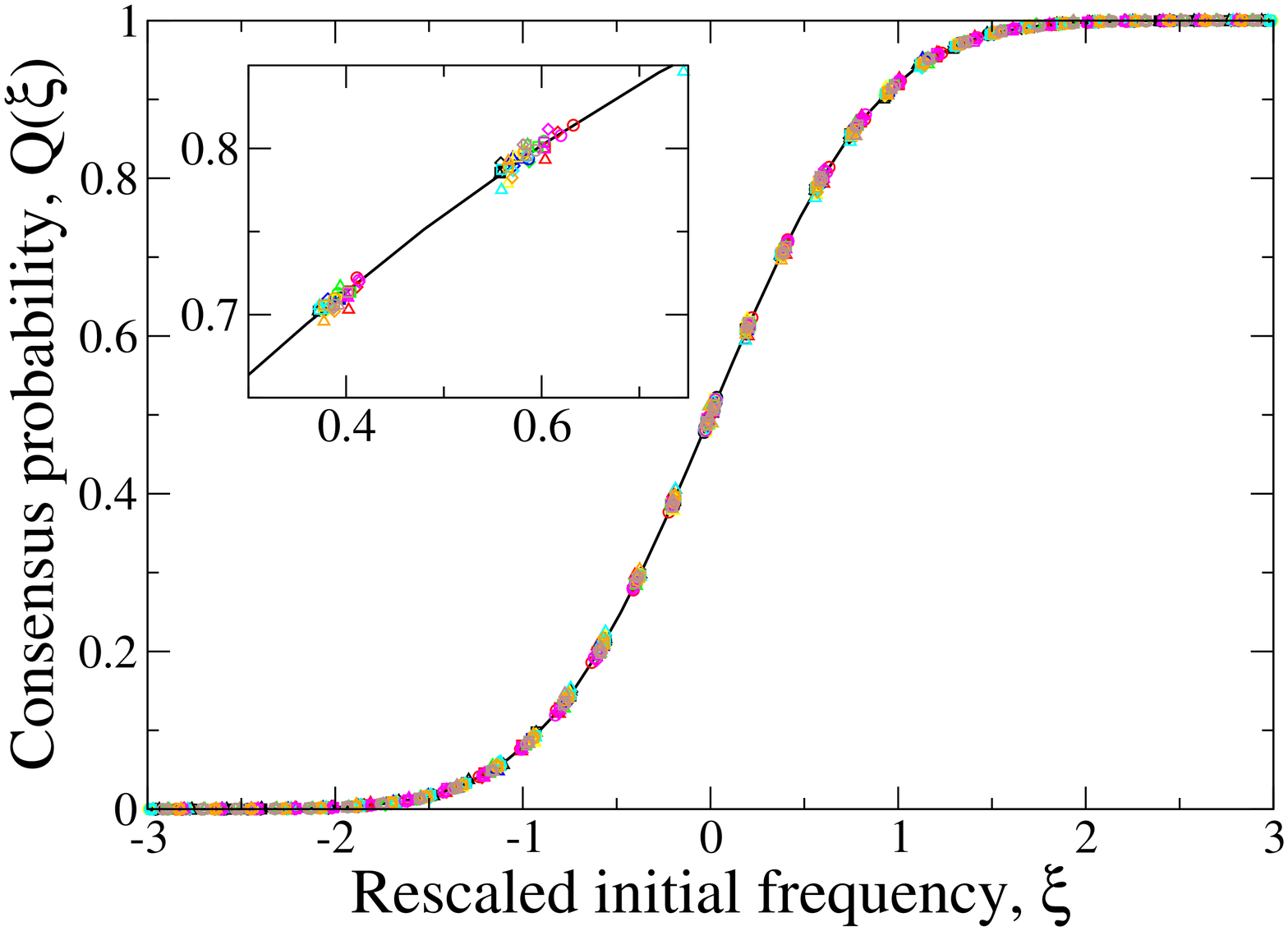}
\hfill
 \includegraphics[width=0.495\linewidth]{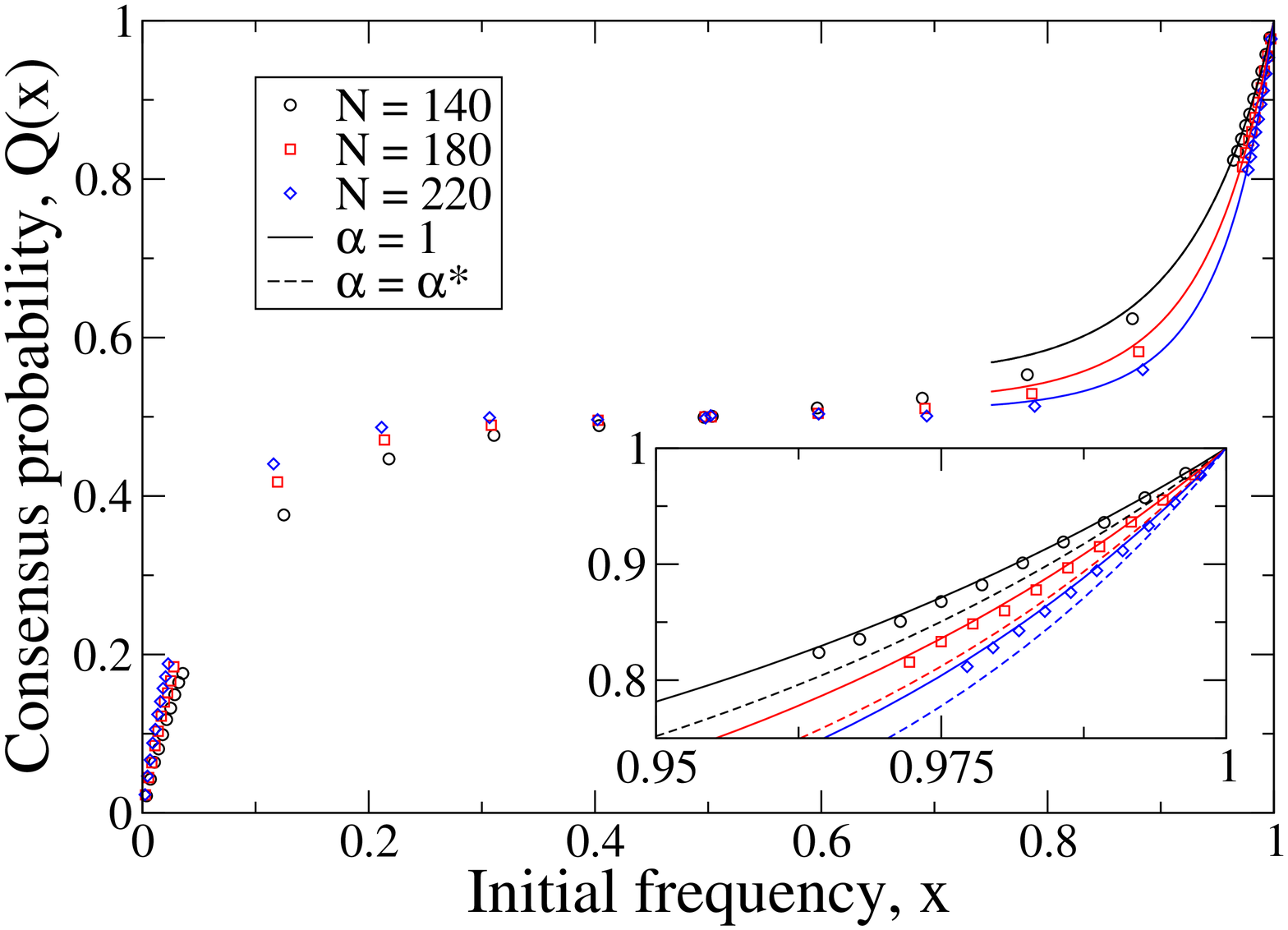}
 \caption{\label{simpleQ} Consensus probability $Q$ as a function of initial token frequency $x$ within the simplified dynamics. Points show results from direct Monte Carlo simulation of the simplified dynamics for a range of different system sizes and choices for the bias and copy parameters; curves show analytical predictions. (a) In the rapid-consensus phase, $\mu>0$, data are expected to collapse onto an error function after change of variable $\xi = \sqrt{\beta/(1+\sigma)}(x-\frac{1}{2})$. Each cloud of points contains results from 36 different simulation conditions (see text), and the two parameters $\alpha$ and $\sigma$ are as given by (\ref{alpha}) and (\ref{sigma}). Errors are approximately the same size as symbol sizes (slightly larger in the inset) and have been omitted for clarity. (b) In the delayed-consensus phase, $\mu<0$, the consensus probability is predicted to approach $\frac{1}{2}$ in the central region, with deviations in boundary regions of size of order $1/N$. Solid lines show a fit to the predicted boundary behaviour (\ref{ynegasymp}) with the parameter setting $\alpha=2x_{AB}^\ast$; dashed lines the prediction with $\alpha=1$.}
\end{figure}

In the regime $\mu>0$, the consensus probability as a function of the initial $A$ token frequency, $x$, is predicted to follow an error-function curve, ${\rm erf}\xi$, in the rescaled variable $\xi = \sqrt{\beta/(1+\sigma)}(x-\frac{1}{2})$. This prediction is easily tested by plotting the fraction of realisations of the dynamics starting from given token frequency $x$ that reach the boundary $x=1$ as a function of $\xi$ for different combinations of model parameters. Fig.~\ref{simpleQ}(a) shows these data for a range of combinations of $b$ and $c$ corresponding to $\mu=\frac{1}{2}(b+c) = \{ \frac{1}{2}, \frac{3}{4}, 1 \}$ and system sizes $N=\{250,500,1000,2000\}$. Agreement with the error function, shown as a solid line, is good.

Meanwhile, when $\mu<0$, the theory predicts that $Q(x) \to \frac{1}{2}$ except in boundary regions of size $~\frac{1}{N}$, where one would expect to see the exponential decays given by (\ref{ynegasymp}). The data shown in Fig.~\ref{simpleQ}(b) for the case $\mu=-0.025$ (obtained through the particular parameter combination $b=-0.65$, $c=0.6$; all combinations give similar results) are consistent with these expectations \emph{as long as} we take $\alpha=1$, rather than $\alpha=2x_{AB}^\ast$, in Eq.~(\ref{ynegasymp}). This is despite the fact that the parameter $\alpha$ enters into the simulation algorithm, and was set to the latter value. One possible explanation for the better fit with $\alpha=1$ is that the consensus probability near the boundaries region is dominated by the dynamics in those regions. There, as observed in Section~\ref{simple}, it was necessary to artificially set the agent frequency $x_{AA}$ to zero should the token frequency $x_A$ become sufficiently small that the approximation $x_{AB}=2\alpha x_A(1-x_A)$ demand a negative frequency of $AA$ agents. This is then equivalent to insisting that $x_{AB}=2x_A$, i.e., a value of $\alpha=1$ in these boundary regions.

\begin{figure}
\begin{center}
 \includegraphics[width=0.8\linewidth]{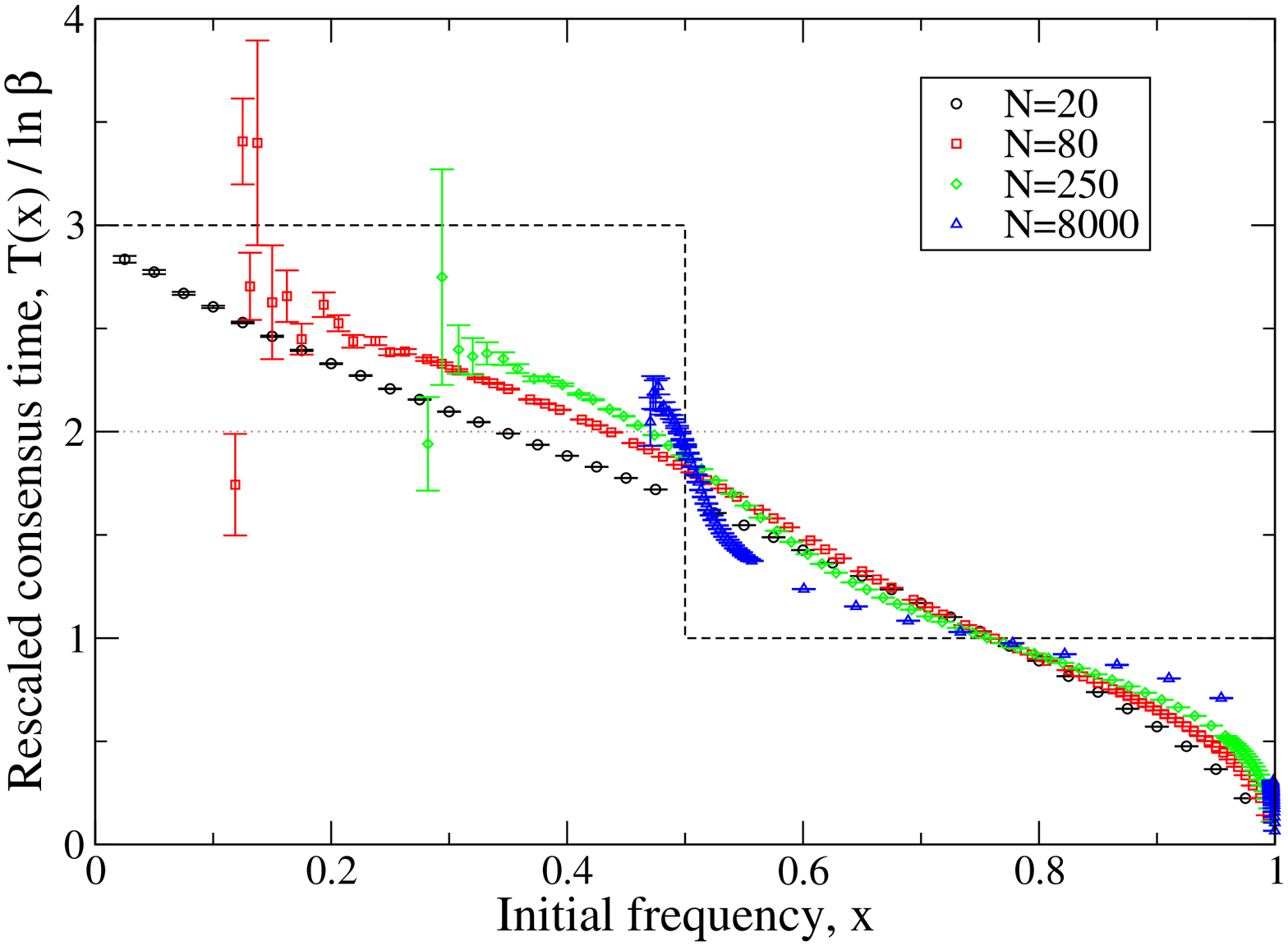}
\end{center}
 \caption{\label{simpleTlnbeta} Leading large-$N$ behaviour of the mean consensus time $T(x)$. The dashed line shows the step-function onto which $T(x)/\ln\beta$ should converge as $N\to\infty$. The points show data from Monte Carlo simulations with $b=c=\frac{1}{2}$ and various $N$. Each point is an average over $10^8/N$ realisations of the dynamics from the initial condition; the error bar shows the standard error on this mean, and is larger for $x<\frac{1}{2}$ because fewer realisations reach consensus on $A$ when it is initially the minority variant.}
\end{figure}

We now examine numerical data for the mean consensus time in the simplified dynamics. The results (\ref{Ts}) indicate that when $\mu>0$, $T(x)$ should converge to a step function with height $3\ln \beta$ at $x<\frac{1}{2}$ and $\ln \beta$ at $x>\frac{1}{2}$ as $N\to\infty$. Different choices of the bias and copy parameters give very similar results: we show as a representative example the case $b=c=\frac{1}{2}$ in Fig.~\ref{simpleTlnbeta}. The data suggest that this convergence will be achieved in the region $x>\frac{1}{2}$, albeit slowly. Even for the largest system simulated, $\ln \beta$ is smaller than $10$, and therefore we expect the $O(1)$ corrections in (\ref{Ts}) still to be significant. In the region $x<\frac{1}{2}$, the situation is less clear. As we have seen, the probability of reaching consensus on $A$ in the regime $x\ll\frac{1}{2}$, for which the consensus time behaviour has been calculated, vanishes exponentially with system size. Therefore it is difficult to sample the large-$N$ behaviour in the stochastic simulations and one cannot state with a high degree of confidence that the simulations reproduce the predicted step-function form. The fact that at the midpoint $x=\frac{1}{2}$, $T(x)$ appears to be converging onto $2\ln \beta$ is perhaps suggestive that this asymptote might eventually be realised, given a sufficient quantity of data for large system sizes.

\begin{figure}
 \includegraphics[width=0.495\linewidth]{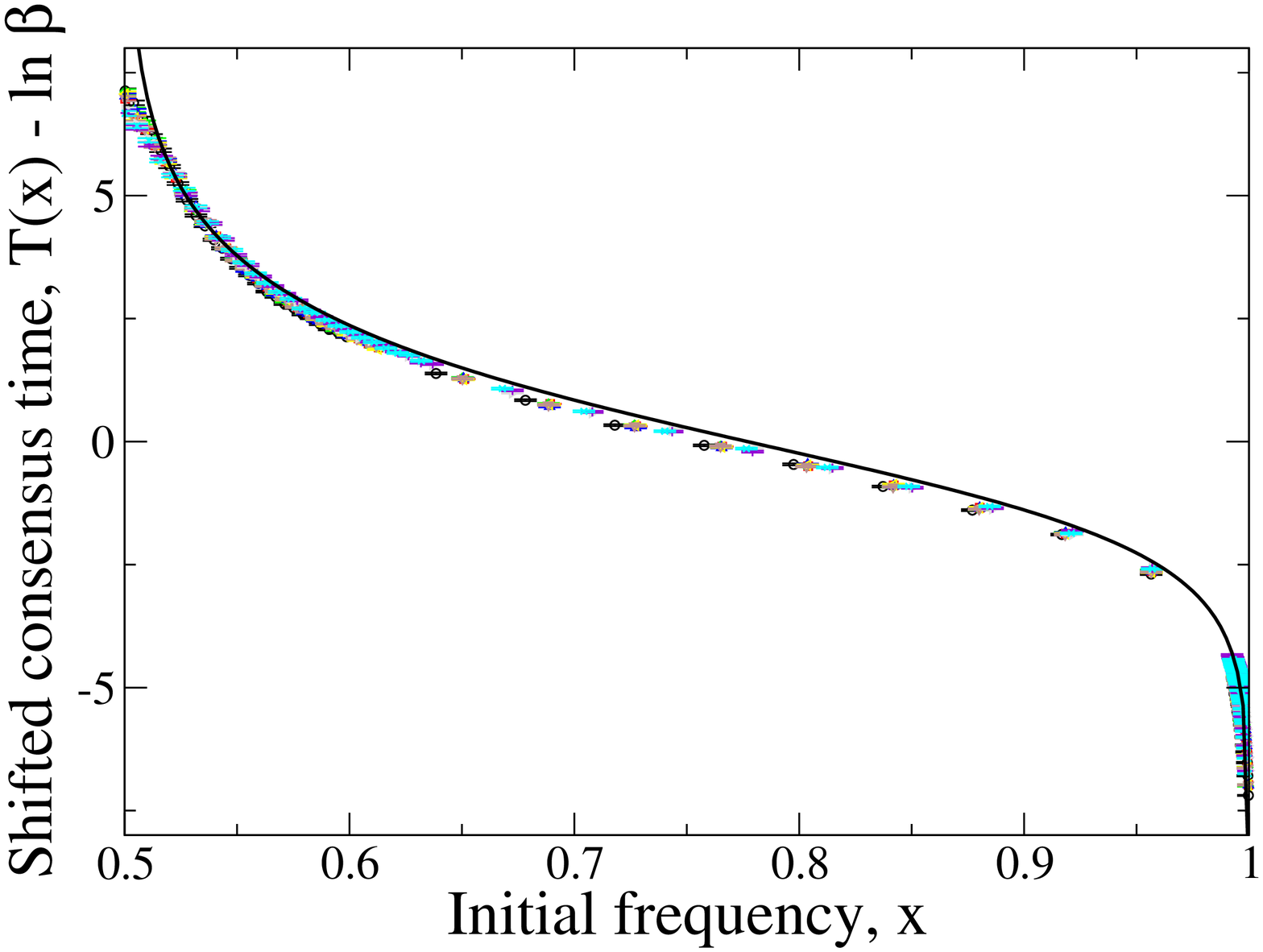}
\hfill
 \includegraphics[width=0.495\linewidth]{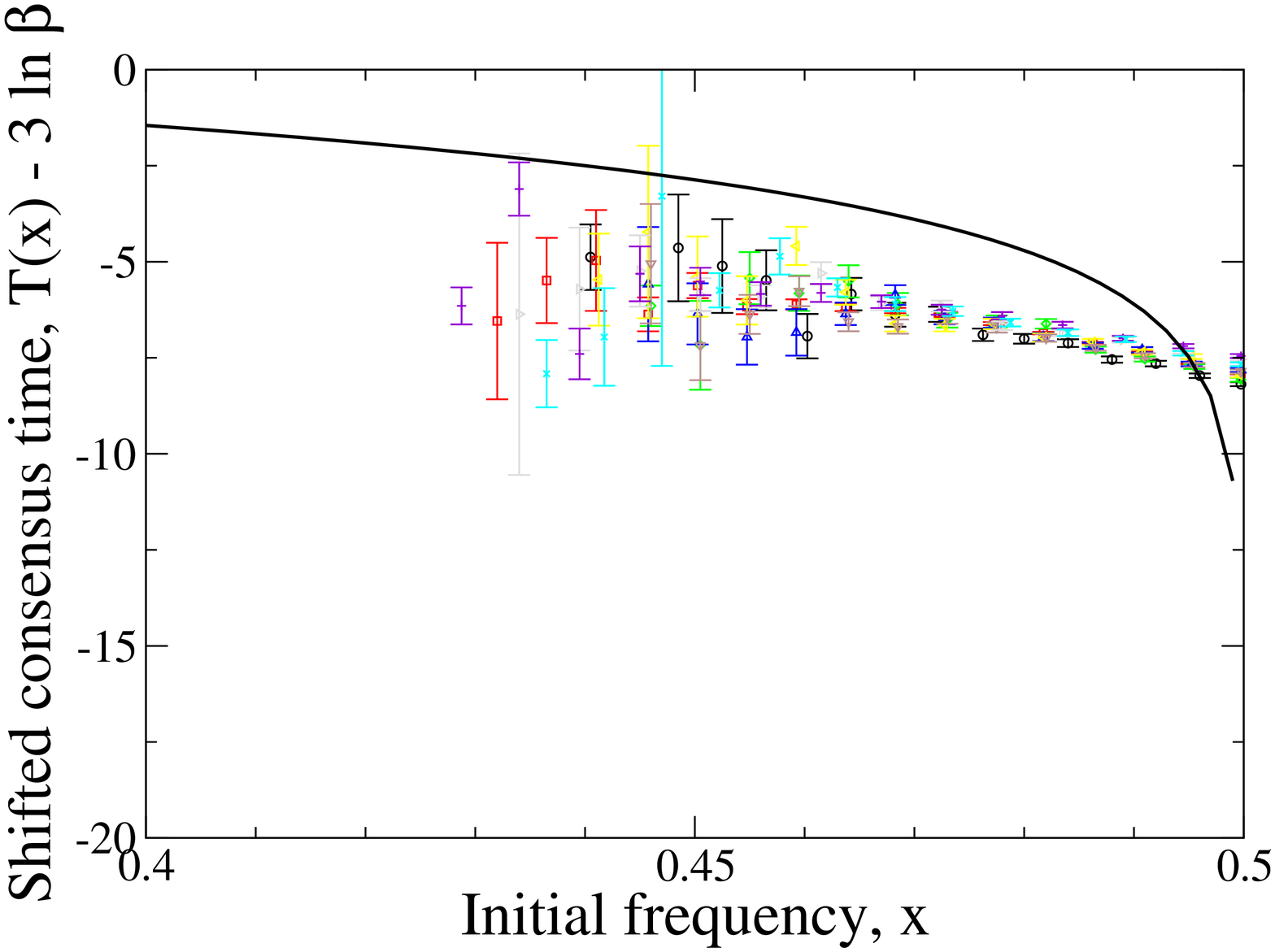}
 \caption{\label{simpleTshift} Difference between the mean consensus time $T(x)$ and the limiting $N\to\infty$ prediction (a) $T(x) \to \ln \beta$ when $x>\frac{1}{2}$ and (b) $T(x) \to 3 \ln \beta$ when $x<\frac{1}{2}$. The solid lines show the functions to which the data should converge as $N\to\infty$, given by (\ref{Ts}). Points show simulation data for various combinations of $b$ and $c$ at $N=2000$. Error bars are as described in the caption to Fig.~\ref{simpleTlnbeta}.}
\end{figure}

We reach similar conclusions when we examine the $O(1)$ contributions to $T(x)$.  In Fig.~\ref{simpleTshift}, we show for a range of $(b,c)$ combinations the difference between $T(x)$ and the leading logarithmic term. The fit to the function $\gamma + \ln [ x(1-x)/(2x-1)^2 ]$ is good in the region $x>\frac{1}{2}$. Again, the data are not necessarily inconsistent with the prediction in the range $x<\frac{1}{2}$, but again, we have the problem that the result (\ref{Ts}) is valid in the regime where $\frac{1}{2}-x \gg 1/\sqrt{\beta}$, which is precisely the regime where the fixation probability is very small. For example, in the plot of Fig.~\ref{simpleTshift}(b), each tick mark on the horizontal axis roughly corresponds to $1/(2\sqrt{\beta})$ for the parameter combinations shown, and we see that the combination $\sqrt{\beta} ( \frac{1}{2} - x)$ rarely exceeds a value of $3$ in practice.

\begin{figure}
\begin{center}
 \includegraphics[width=0.8\linewidth]{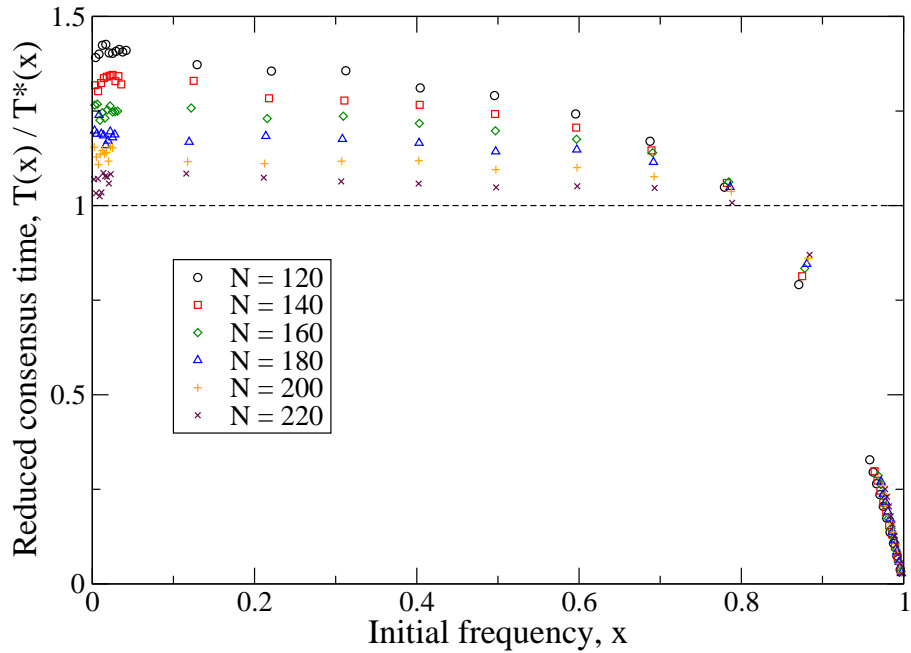}
\end{center}
\caption{\label{simplenT} Consensus time $T(x)$ divided by the prediction $T^*$ given by the right-hand side of (\ref{Txneg}) for the $\mu<0$ regime. The specific parameter choices used were $b=-0.65$, $c=0.6$ (and hence $\mu=-0.025$) with $10^7/N$ repetitions of the dynamics from each initial condition.}
\end{figure}

Data for the consensus time in the delayed-consensus phase are plotted in Fig.~\ref{simplenT}, divided by the prediction (\ref{Txneg}), so that as $N\to\infty$, one would expect data to converge to the line $y=1$ (except within a region of size $~1/N$ near $x=1$). Combinations of model parameters other than those used in the figure ($b=-0.65$, $c=0.6$) show a similar approach to this line, so taken together we would suggest that the predicted convergence is reflected in the numerical data. Certainly we find a more convincing convergence when the parameter $\alpha=2 x_AB^\ast$ (the value used for the figure) than when $\alpha=1$ (data not shown). This is consistent with our earlier observation that this would be the appropriate choice for $\alpha$ in the delayed-consensus regime for quantities dominated by the dynamics in the central region (as opposed to the boundaries). As we saw in Section~\ref{large}, the dominant contribution to consensus time is the time needed to escape the potential well that is centred at $x=\frac{1}{2}$ and whose depth is large compared to the scale of the fluctuations.

To summarise, analytical results for the consensus probability obtained in the large-$N$ limit appear to give a good fit to simulation data for finite systems in both the positive and negative $\mu$ regimes, albeit with a modified choice of $\alpha=1$ to describe the boundary behaviour in the $\mu<0$ regime. The predicted mean consensus time behaviour is reproduced for $\mu>0$ and $x \ll \frac{1}{2}$. When $\mu>0$ and $x \ll \frac{1}{2}$, consensus on the $A$ variant is sufficiently rare that it is difficult to discern whether the predicted behaviour is observed; and when $\mu<0$, the finite-size effects seem still to be strong over the range of $N$ within which consensus occurs quickly enough to be seen in simulation. One possible way to strengthen these conclusions would be to use specialised techniques for sampling rare events (such as forward flux sampling \cite{all05}), which we leave as a possibility for future work.

\subsection{Correspondence between the simplified and full dynamics}

The results for the consensus probability and mean consensus time are given in terms of the initial overall frequency $x$ of $A$ tokens. In the full dynamics, a single value of $x$ may be realised through various relative numbers of $AA$ and $AB$ agents. To test whether the predictions based on the simplified dynamics extend to the full dynamics, we must, for any given value of $x$, choose several compatible initial conditions. We ran simulations with initial $x_{AB} = \{0, \frac{1}{2} x_{\rm max}, x_{\rm max} \}$ where $x_{\rm max}$ is the maximum possible fraction of $AB$ agents that can be realised for given $x$. For $x>\frac{1}{2}$, $x_{\rm max} = 2 (1-x)$. Once $x$ and $x_{AB}$ are specified, $x_{AA}$ is given by $x - \frac{1}{2} x_{AB}$, and $x_{BB}$ through the normalisation $x_{AA}+x_{AB}+x_{BB} = 1$.

\begin{figure}
 \includegraphics[width=0.495\linewidth]{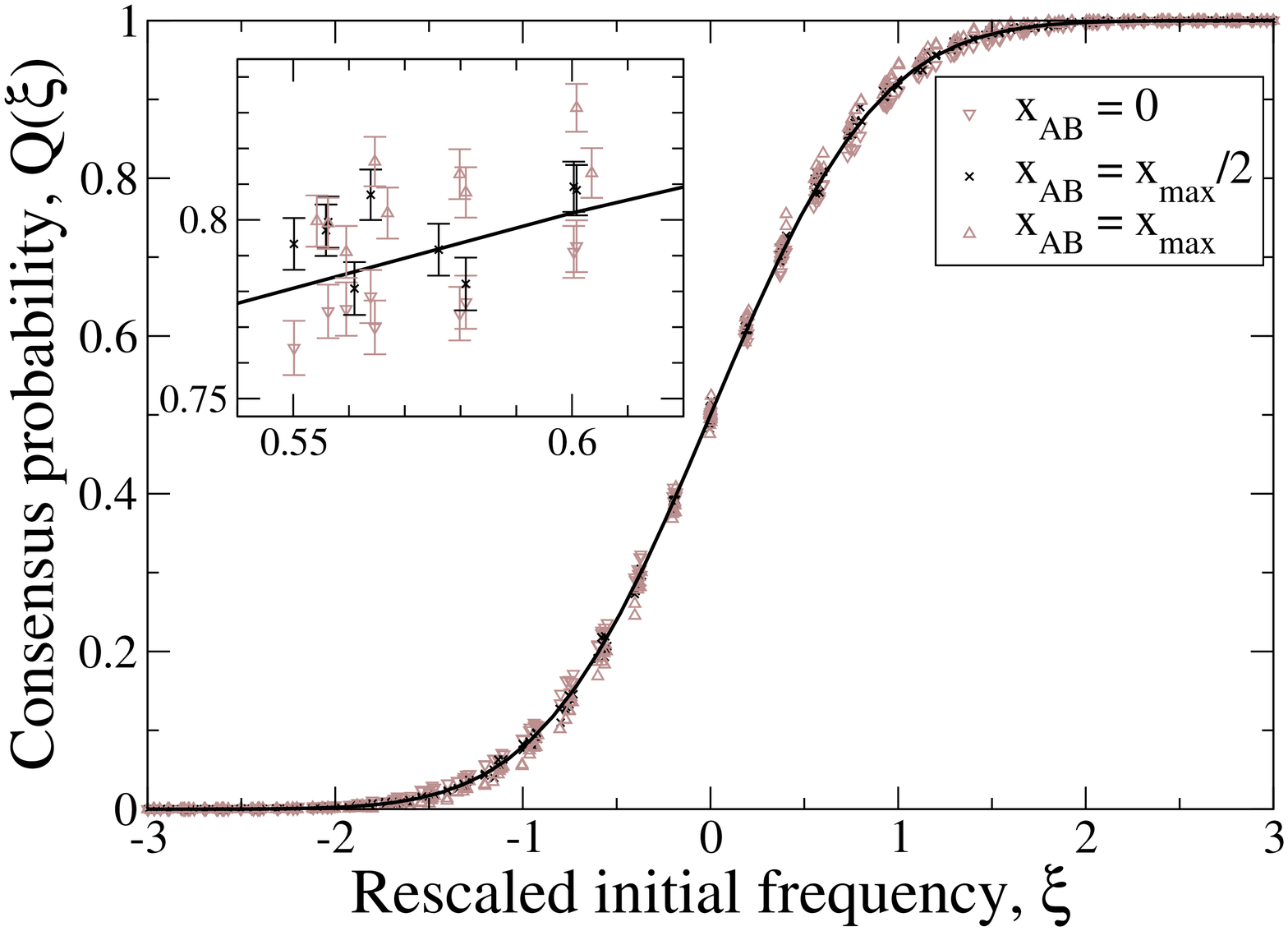}
 \includegraphics[width=0.495\linewidth]{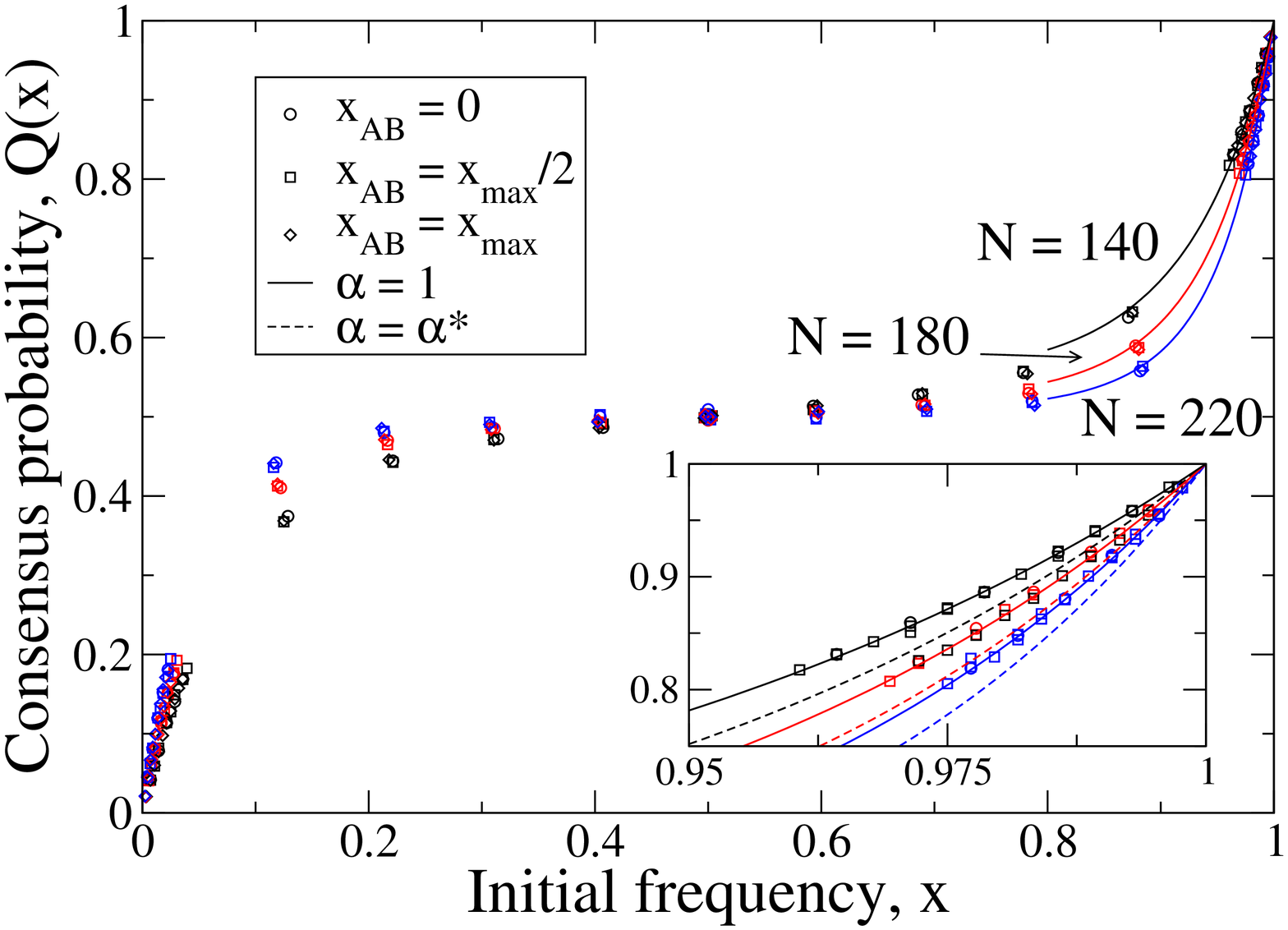}
 \caption{\label{fullQ} Consensus probability $Q(x)$ within the full dynamics. In both figures, different symbols relate to different initial fractions of inconsistent ($AB$) agents for a given value of $x$. Errors are approximately the same size as the symbols in the main figures. (a) In the rapid-consensus regime $\mu>0$, $x$ is rescaled via $\xi = \sqrt{\beta/(1+\sigma)}(x-\frac{1}{2})$ as in Fig.~\ref{simpleQ}. Points show results from Monte Carlo simulations within communities of $N=16000$ agents, each initial condition repeated $3125$ times.  The solid curve is the prediction from the simplified dynamics. The values of $\alpha$ and $\sigma$ used to transform from $x$ to $\xi$ are as given by (\ref{alpha}) and (\ref{sigma}). (b) The corresponding data for the delayed-consensus phase, $\mu<0$. As in Fig.~\ref{simpleQ}, solid lines show a fit to the predicted boundary behaviour (\ref{ynegasymp}) with the parameter setting $\alpha=2x_{AB}^\ast$; dashed lines the prediction with $\alpha=1$.}
\end{figure}

The first test of the predictions is whether the error-function form of the consensus probability $Q(x)$ applies when $\mu>0$. Fig.~\ref{fullQ} shows simulation data obtained for various $b$ and $c$ and, for each $x$, the three initial conditions just described. The data fit the error function reasonably well, but closer inspection shows that whilst the fit is best for the set of data that have the initial $x_{AB}=x_{\rm max}/2$, those for $x_{AB}=0$ and $x_{AB}=x_{\rm max}$ consistently lie respectively below and above the predicted curve in the range $x>\frac{1}{2}$. Presumably these discrepancies are due to the fact that these initial conditions most strongly violate the assumption that $x_{AB} = 2\alpha x_A(1-x_A)$, invoked to obtain the simplified dynamics. A better prediction for the function $Q(x)$ generated by the full dynamics would therefore somehow need to take the initial number of inconsistent ($AB$) agents into account.  We remark that transforming these data with $\alpha=1$, rather than $\alpha=2x_{AB}^\ast$, yields a worse overall fit to the error function. This is in accordance with our earlier observation that quantities dominated by $x\approx\frac{1}{2}$ (which, in the analysis, is where the error-function form comes from) are better predicted by the choice $\alpha=2x_{AB}^\ast$. In the delayed-consensus phase, $\mu<0$, we find---as with the simplified dynamics---a better fit of (\ref{ynegasymp}) to the data near the boundaries is achieved by taking $\alpha=1$, as is shown by Fig.~\ref{fullQ}(b).

\begin{figure}
 \includegraphics[width=0.495\linewidth]{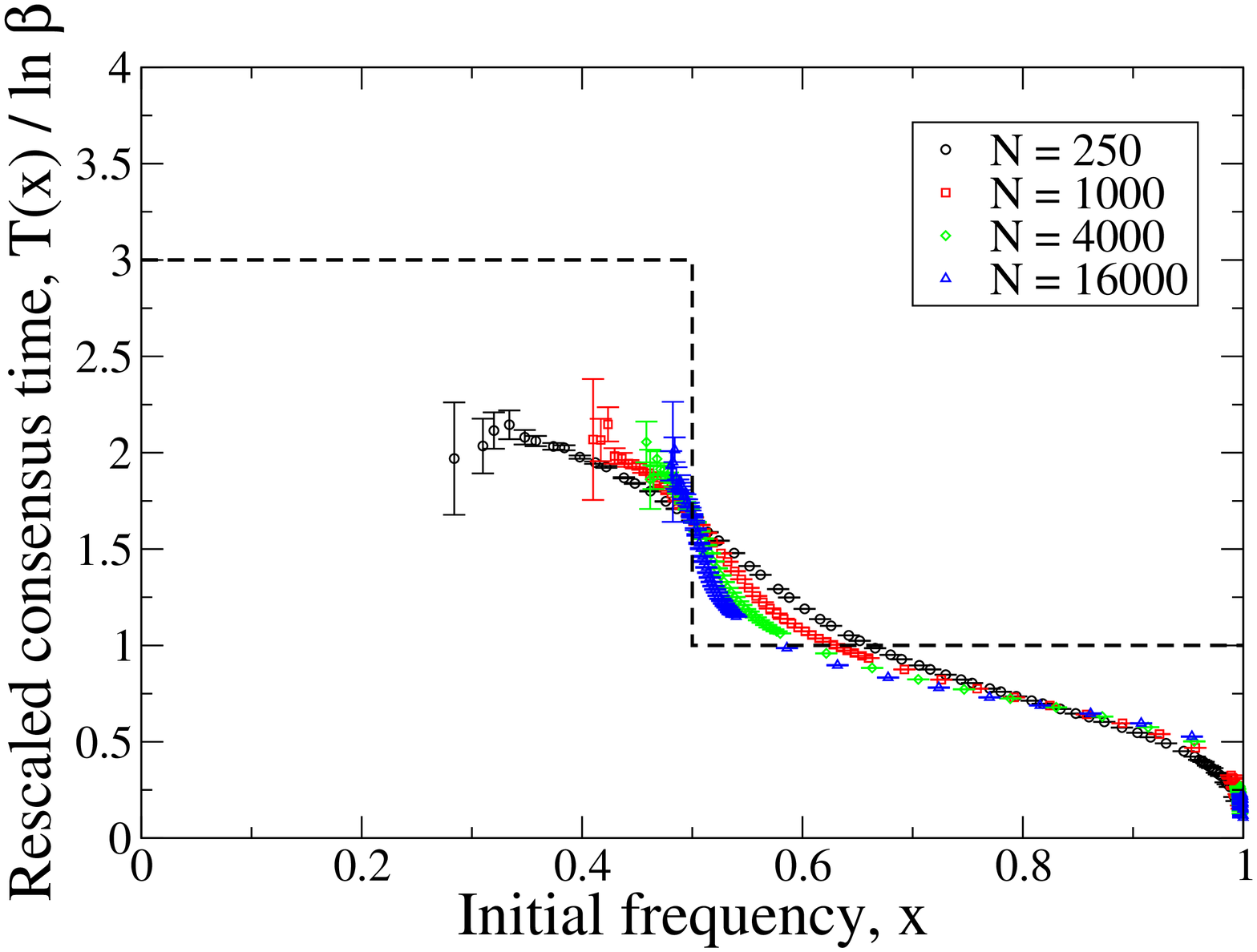}
 \hfill
 \includegraphics[width=0.495\linewidth]{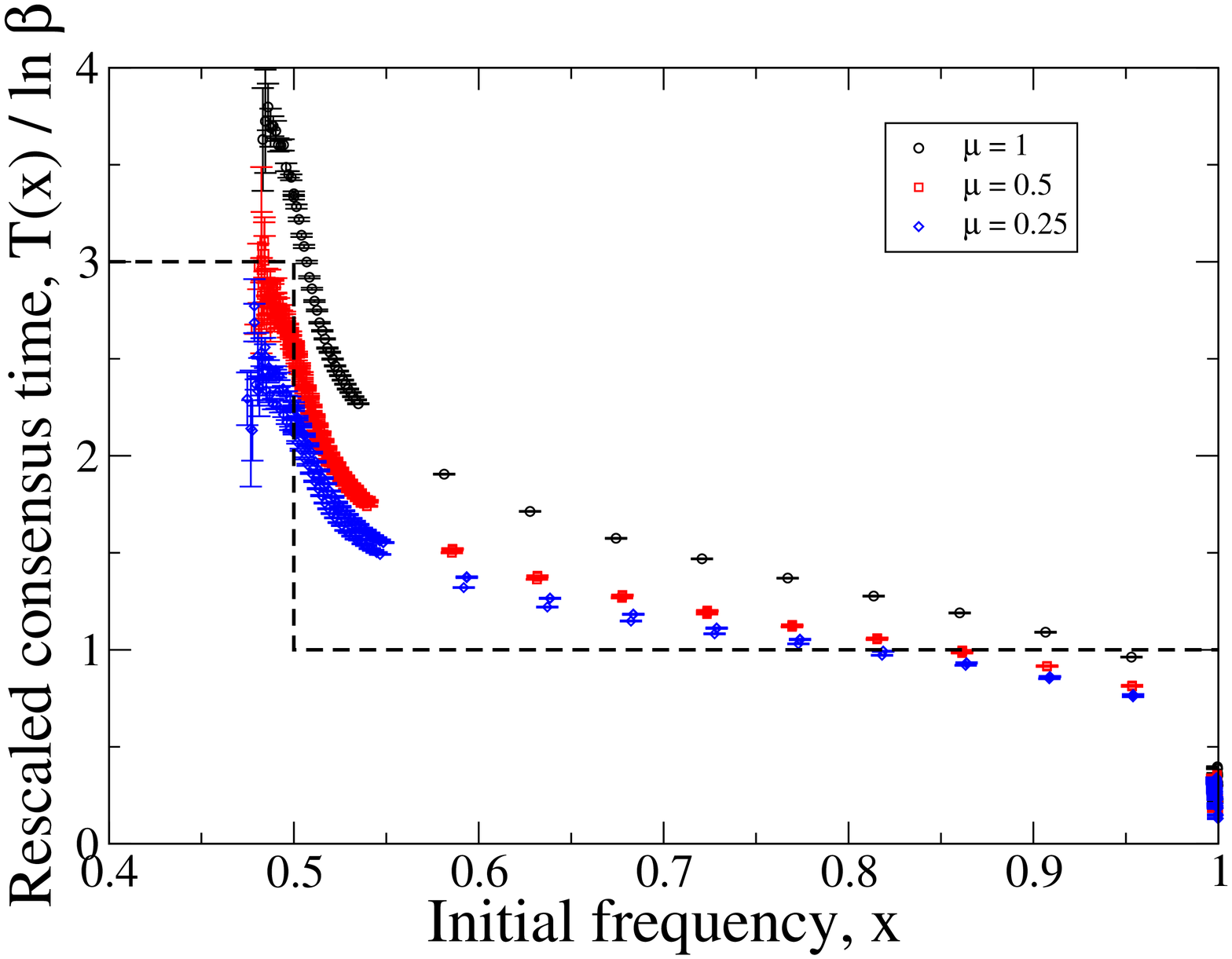}
 \caption{\label{fullTlnbeta} (a) As Figure~\ref{simpleTlnbeta} except that the points were obtained from simulations of the full hybrid model dynamics, not the simplified dynamics. The condition $b=c=\frac{1}{2}$ is the same; the initial fraction of inconsistent $AB$ agents is $x_{\rm AB}=1-x$; and each initial condition was repeated $5\times 10^7/N$ times. (b) Data from the same simulations, but at fixed $N=16000$ and various combinations of $b$ and $c$.}
\end{figure}

We now examine how well predictions for the mean consensus time from the simplified dynamics carry over to the full dynamics. Whilst data for the simplified dynamics were not inconsistent with asymptotic convergence of the function $T(x)/\ln\beta$ onto the predicted step function, those for the full dynamics are far less convincing, as shown by Fig.~\ref{fullTlnbeta}(a). Taking the point at which the curves intersect as a guide to the height of the step in the region $x>\frac{1}{2}$, we see that the prediction (\ref{Ts}) overestimates this by about $30\%$. It is natural to hypothesise that the discrepancy is due to the initial condition being incompatible with the restriction $x_{AB} = 2 \alpha x_A(1-x_A)$ used to formulate the simplified dynamics. However, we find that the simulated consensus time data (not shown) are largely insensitive the initial value of $x_{AB}$. Instead, we find that different combinations of $b$ and $c$ that have the same value of $\mu=\frac{b+c}{2}$ give roughly similar consensus times, and that the main variation is between different values of $\mu$: see Fig.~\ref{fullTlnbeta}(b).

\begin{figure}
\begin{center}
 \includegraphics[width=0.8\linewidth]{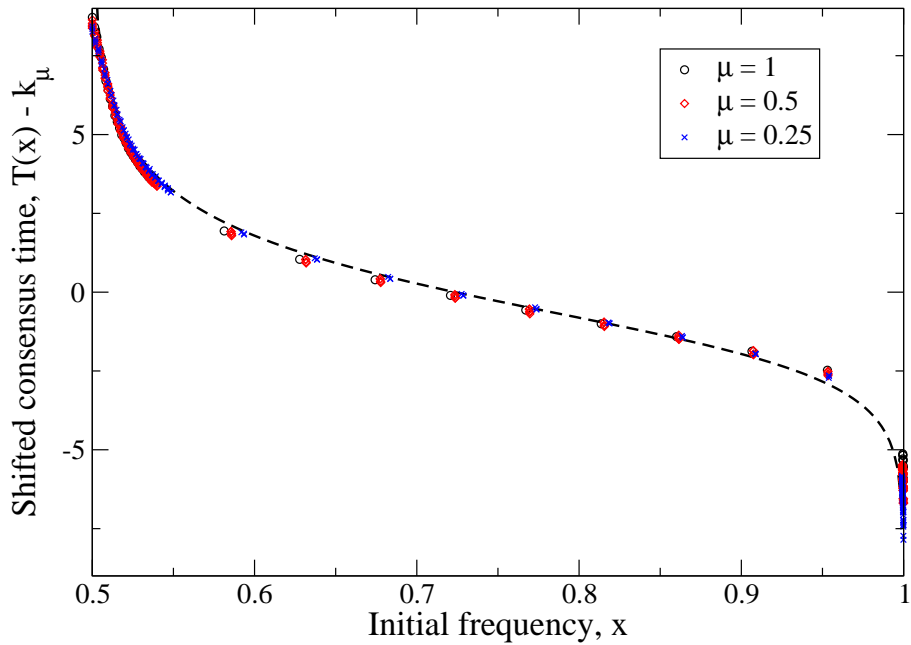}
\end{center}
 \caption{\label{fullTshift} Same consensus time data within the full dynamics as in Fig.~\ref{fullTlnbeta}(b), but instead of being scaled by a factor $\ln\beta$, shifted by an amount $k_{\mu}$ to obtain a roughly common curve (as judged by the eye). The dashed line is the function $\ln x(1-x)/(2x-1)^2$ predicted by the deterministic part of the simplified dynamics.}
\end{figure}

What is remarkable is that the \emph{shape} of the consensus time function $T(x)$ obtained by simulation is consistent with the analytical expression $\gamma+\ln x(1-x)/(2x-1)^2$ predicted by the simplified dynamics. To see this, we plot $T(x)$ for different $\mu$, and shift each data set by a $\mu$-dependent constant $k_{\mu}$: see Fig.~\ref{fullTshift}. We have also observed that this constant is independent of the community size $N$ if one makes the mean-field approximation $\alpha=1$ (rather than $\alpha=2x_{AB}^\ast$). Again this could be a consequence of the leading $\ln\beta$ contribution to the consensus time originating from behaviour near the boundary, specifically, the linear vanishing of both the deterministic and the diffusion terms in the Fokker-Planck equation (\ref{fpe}).

\begin{figure}
\begin{center}
 \includegraphics[width=0.8\linewidth]{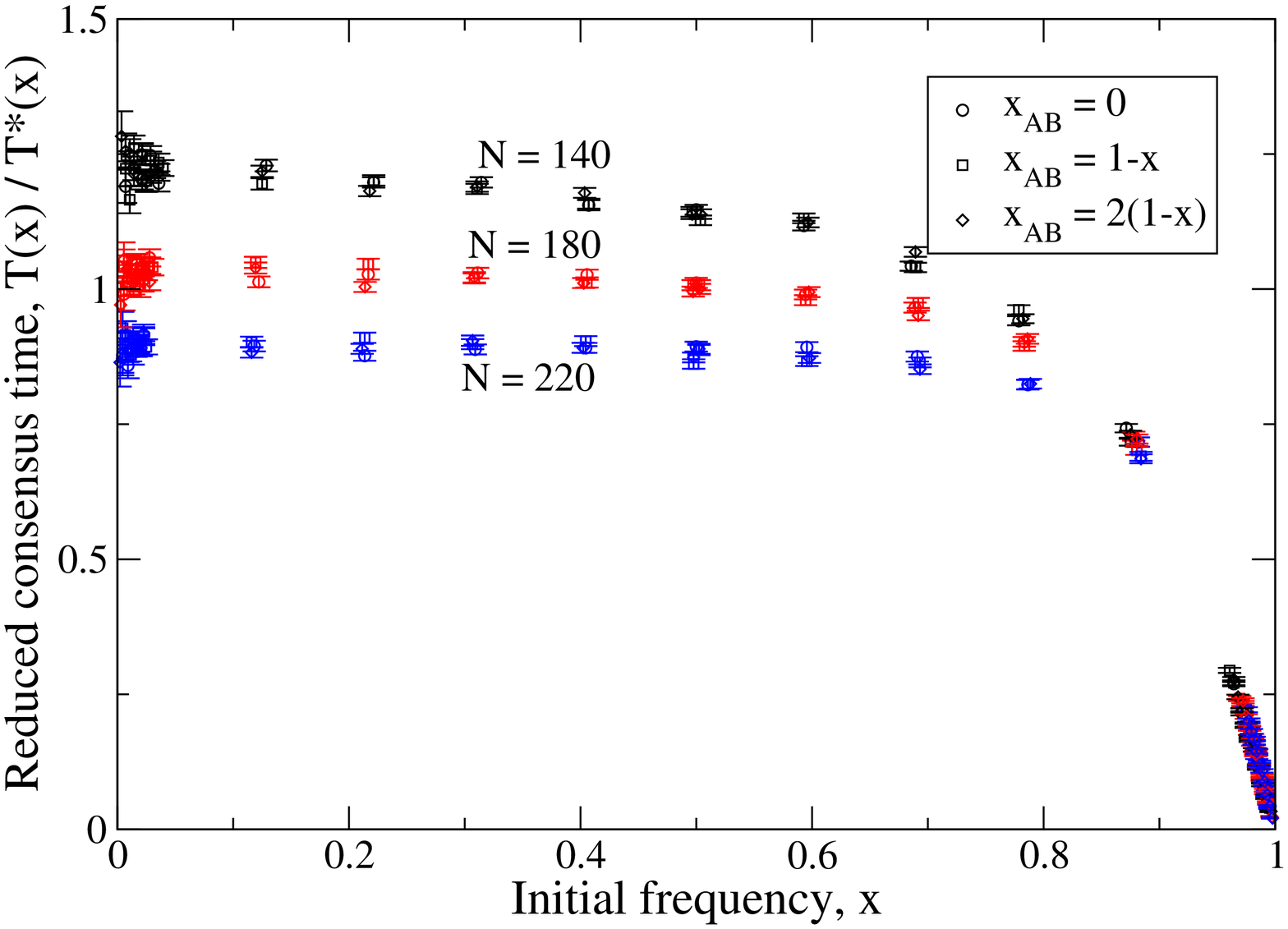}
\end{center}
\caption{\label{fullnT} Measurement of the consensus time $T(x)$ within the full dynamics for $b=-0.65$, $c=0.6$ and hence $\mu=-0.025$. As in Fig.~\ref{simplenT}, $T(x)$ has been divided by the prediction $T^*$ given by the right-hand side of (\ref{Txneg}). The different symbols correspond to the three different initial conditions, and different colours to different system sizes.}
\end{figure}

Similar trends are displayed by data obtained for the delayed-consensus phase: the true consensus time seems to fall below that predicted by the simplified dynamics and and is insensitive to the initial condition, as can be seen from Fig.~\ref{fullnT} for the specific parameter combination $b=-0.65$ and $c=0.6$. Other parameter choices do not necessarily show the undershoot evident in Fig.~\ref{fullnT}: for example, with $b=-0.05, c=0$ (not shown), convergence to the consensus time given by (\ref{Txneg}) is more convincing. Taking all the data for different combinations of $b$ and $c$ together it is not clear whether taking (\ref{Txneg}) with $\alpha=1$, rather than $\alpha=2x_{AB}^\ast$, gives overall a better fit. We therefore conclude that although the data for the full dynamics are consistent with an exponential growth in the consensus time with the community size $N$, predictions from the simplified theory cannot be confirmed or rejected without access to data for larger system sizes.

\begin{figure}
 \includegraphics[width=0.495\linewidth]{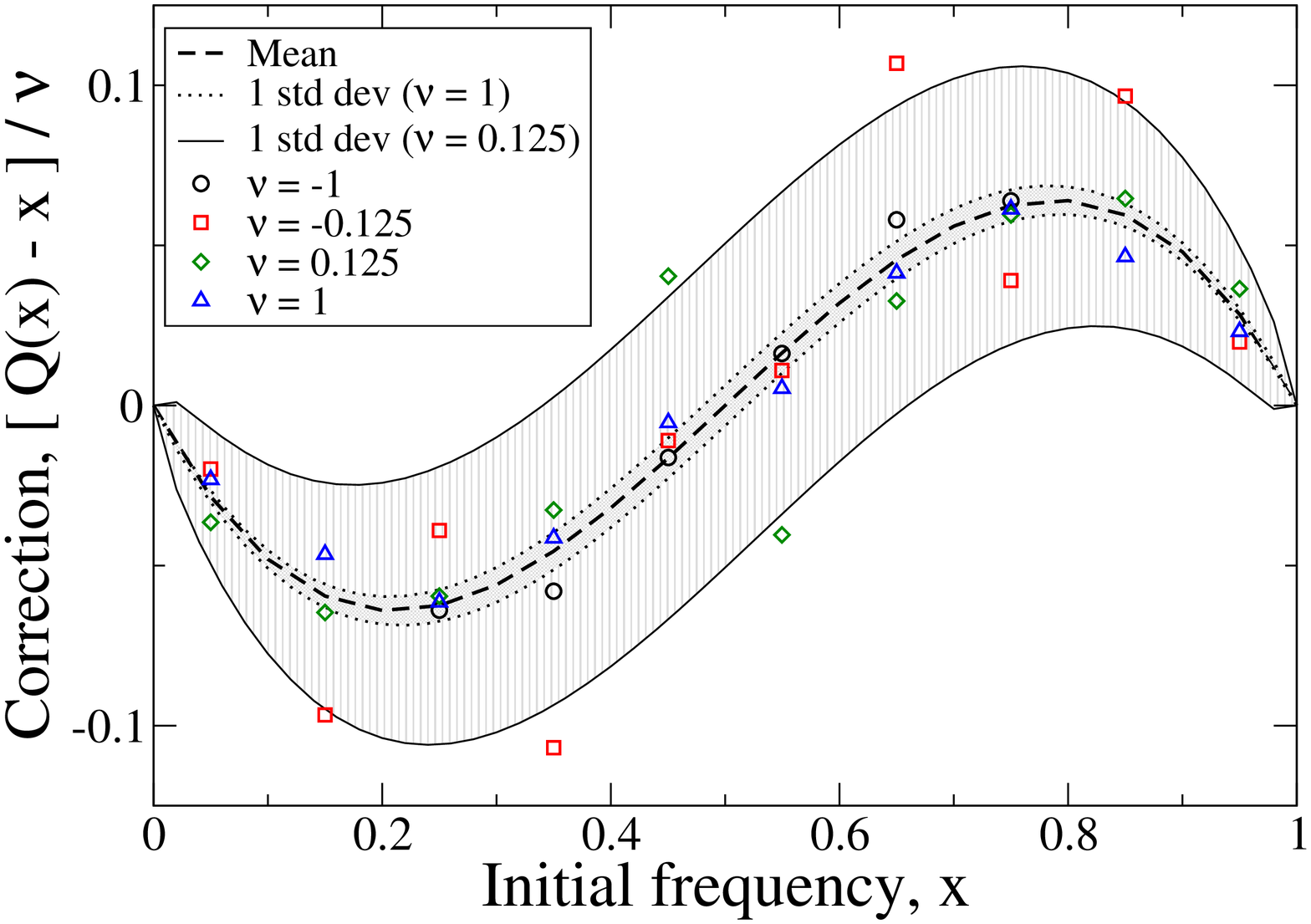}
 \hfill
 \includegraphics[width=0.495\linewidth]{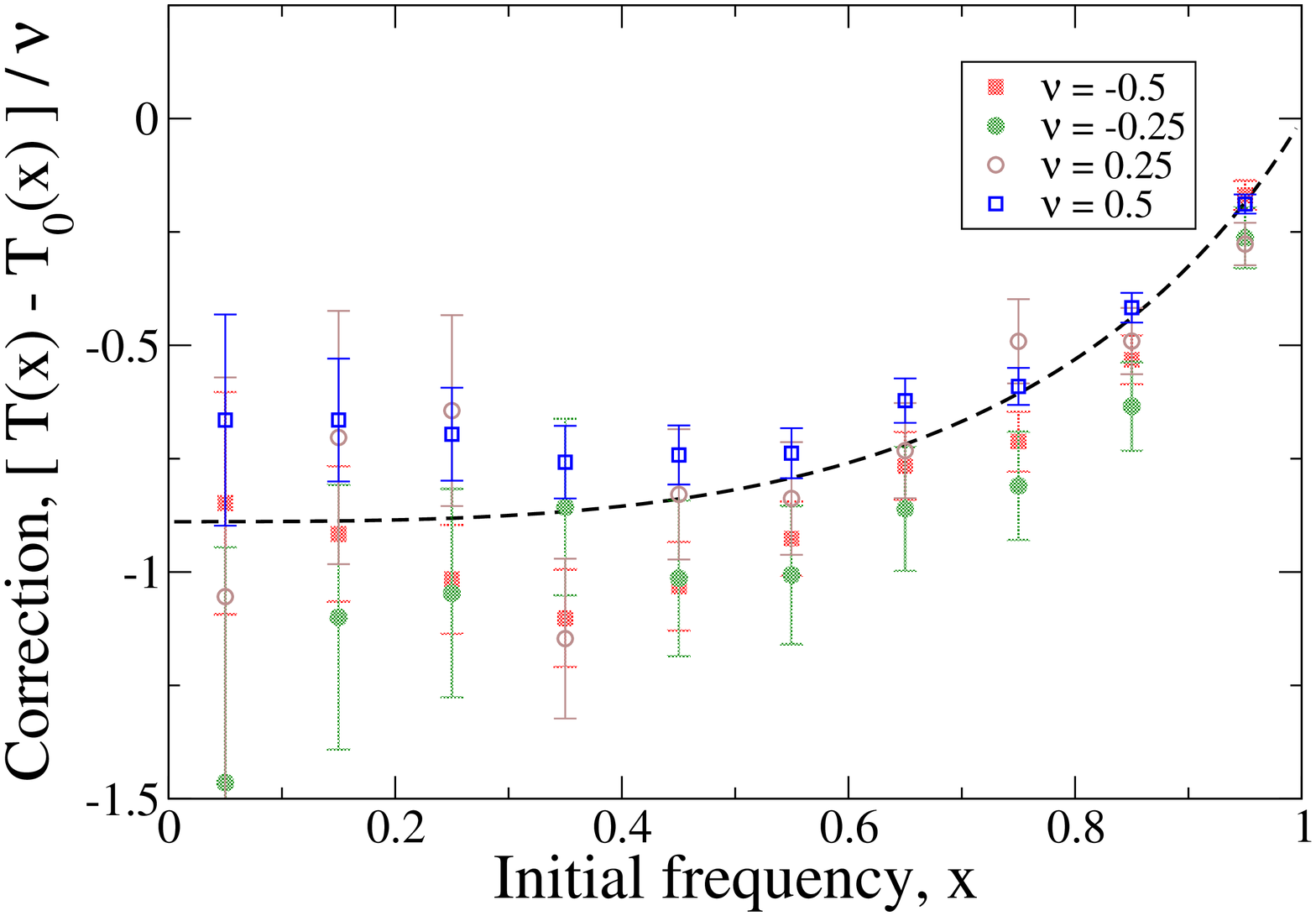}
\caption{\label{fulld} (a) Deviation of the function $Q(x)$ from the $\nu=0$ limit within the full dynamics in a community of $N=1600$ speakers and $6250$ repetitions from each initial condition.  Dashed line shows the expectation value for each point; the inner and outer intervals one standard deviation around this for $|\nu|=1$ and $|\nu|=0.125$ respectively. (b) The analogous data for the consensus time function $T(x)$, this time with standard error on each sampled mean displayed explicitly.}
\end{figure}

To finish, we briefly examine the crossover regime, $|\mu| \sim 1/N$.  In Section~\ref{large}, we found an expansion around the pure Voter Model behaviour as a series in a parameter $\nu$ defined by (\ref{nulim}) for the case where both $b$ and $c$ vanish with $N$ as $1/N$ (Case I). To test the validity of this expansion within the full dynamics, we plot the difference between the measured consensus probability and time from the Voter Model values---i.e., the leading order terms in (\ref{Qd}) and (\ref{Td})---divided by $\nu$ and compare with the coefficient of the first-order terms. This we did for $N=1600$ and a range of $\nu$, both positive and negative. For the positive $\nu$ values, we took $b=c=\nu/N$; for the negative values, $b=-3\nu/N$ and $c=\nu/N$. The desirable range of $\nu$ is such that it is sufficiently small for second order effects to be negligible, but sufficiently large that deviations of $Q(x)$ from the $\nu=0$ limit $Q(x)=x$ are not swamped by the noise. To this end, we plot in Fig.~\ref{fulld}(a) the first-order term in (\ref{Qd}), the expected mean value of the sampled quantity, and around it the expected standard deviation of the data around this mean corresponding to the two different values of $|\nu|$ that are displayed.  The data are consistent with these intervals. The consensus time data, Fig.~\ref{fulld}(b), when errors are taken into account are also broadly consistent with the prediction from the simplified dynamics but, in common with the other data discussed in this section, to a lesser degree than the consensus probability data.

In summary, then, we find that the simplified dynamics acts as a good proxy for the full dynamics as far as the consensus probability is concerned, and that the additional freedom in the distribution of $A$ tokens between consistent $AA$ agents and inconsistent $AB$ agents has a fairly small effect on the consensus probability function. The consensus time, meanwhile, seems to be insensitive to different ways of setting up an initial condition with overall $A$ token frequency $x$, but depend on the parameters $b$ and $c$ in a way that is not predicted by the simplified dynamics. This is particularly evident in the regime $\mu>0$ and $x>\frac{1}{2}$, where---up to a $\mu$-dependent shift---the shape of the consensus time function is well-described by the function arising from the simplified dynamics. As with the simplified dynamics, we find that quantities governed by behaviour near the boundaries are better fit by taking $\alpha=1$, and that finite-system and -sample size effects hamper our ability to make strong conclusions in the regimes $\mu>0, x<\frac{1}{2}$ and $\mu<0$. Meanwhile, the behaviour in the crossover regime of the full dynamics appears to be well-captured by the perturbative expansion performed within the simplified dynamics.

\section{Discussion and outlook}
\label{discuss}

In this work, our aim has been to obtain a generic understanding of consensus formation in multi-agent systems by bringing together hitherto disparate statistical-mechanical models of language dynamics. By restricting to the empirically-relevant case of two variant forms that are competing to become the single convention shared by all members of a community, we have been able to unify the dynamics of the much-studied Voter Model and its relatives (one of which is the Utterance Selection Model \cite{bax06}) with that of the Naming Game \cite{bar06}. Our analysis shows that the specific implementation of maximising behaviour employed in the Naming Game leads to an effective repulsion in the frequencies of the $A$ and $B$ variants in the community mediated by what we have called inconsistent agents, i.e., those who use both variants equally often. Certain parameter choices correspond to an `anti-maximisation' behaviour, in which agents are reluctant to abandon the notion that either variant may equally well represent the target meaning. A similar, but distinct, update rule had previously been implemented in a model due to Baronchelli \textit{et al} \cite{bar07}, and in both the hybrid model family and that described in \cite{bar07} the same generic phase structure is seen. In one phase, consensus on the majority variant is reached, and in the other, both variants coexist with equal frequency. This phase diagram was found to be robust to the addition of noise, which permits consensus even among anti-maximising agents, albeit after a time that grows exponentially with the community size.

The analysis of the stochastic dynamics was achieved by replacing the fraction of inconsistent agents, a stochastic variable, with a deterministic function of the frequency of $A$ variants in the community. Despite the somewhat crude and uncontrolled nature of this approximation, we found from Monte Carlo simulations that the consensus probability from a given initial condition was nevertheless well described, and although the growth of the consensus time with community size (logarithmic and exponential in the rapid and delayed consensus phases, respectively) predicted by the simplified dynamics was seen in simulations, the simplified model does not  provide a precise quantitative description of the consensus time behaviour seen in the full dynamics.

Rather than dwell on the inadequacies of the simplified dynamics as a proxy for the full dynamics, discussed at length in Section~\ref{sim}, we will now instead consider what light the present study sheds on linguistic consensus-formation processes in general. It is interesting to note that the form of drift and diffusion terms in the Fokker-Planck equation (\ref{fpe}) are precisely what one would write down in an abstract formulation following the spirit of Landau free energy theory. That is, they are both the lowest-order polynomial expressions that respect the symmetries and boundary conditions of the problem. For consensus to be an absorbing state, it is necessary for both to vanish when a single variant remains. Furthermore, if we restrict ourselves to neutral theories, i.e., those in which $A$ and $B$ variants are distinguished only by their frequencies, we require that the drift term is antisymmetric around $x=\frac{1}{2}$, and that the diffusion term is symmetric. The expressions $x(1-x)(1-2x)$ and $x(1-x) + \sigma x^2(1-x)^2$ are, indeed, the lowest order terms in a Landau-like expansion with these properties. It is reassuring that these expressions were obtained for concrete models that have been specifically introduced to study language dynamics. Furthermore,
precisely these forms are also found for the distinct class of models discussed in \cite{bar07} (with appropriate choices of the parameters $\mu$ and $\sigma$). Thus we should expect to find for those dynamics too that at a \emph{finite} system size $N$, a cross-over to Voter-Model-like behaviour emerges at a distance of order $1/N$ from the transition point. Given that these different microscopic dynamics lead to the same effective description, and that the two distinct maximising rules in the Naming Game enter into the expressions in a similar way, we anticipate that other concrete models of maximising behaviour are likely to lead to similar macroscopic dynamics, at least in mean-field communities.

The key properties of these expressions that dominate the collective behaviour of the hybrid model are: (i) the presence of a maximum or minimum in the potential defined by Eq.~(\ref{aVb}); and (ii) the manner in which the drift and diffusion terms vanish at the boundaries. A potential maximum leads to the error-function form of the consensus probability in the rapid consensus phase (via a harmonic approximation at the maximum), whilst a minimum ultimately gives rise to the exponential growth with inverse temperature in the time needed to escape it. The vanishing drift term leads to a divergence in the consensus time which is controlled by a cut-off governed by the form of the potential near the boundaries, as demonstrated by Eq.~(\ref{Ixlo}). A leading linear decay of the drift term at the boundaries would appear always to imply a logarithmically-growing consensus time; we anticipate that other leading terms will cause polynomial growth of the consensus time with population size. Meanwhile, a potential containing multiple maxima and minima will likely lead to a model that displays a mixture of the behaviour we have seen here, that is, consensus probability functions that take error-function forms around potential maxima and exponentially growing contributions to consensus times arising from each minimum (along with logarithmic or polynomial contributions from boundary regions). To firm up these speculative conclusions, it would perhaps be worthwhile in the future to investigate more systematically the range of possible emergent behaviour arising from different forms of the drift and diffusion terms, and also to develop a more controlled scheme for reducing a multi-dimensional dynamics to a single coordinate, within which such concepts as a potential take on a straightforward interpretation.

A number of studies have been devoted to establishing connections between the Voter and Ising models \cite{oli93,dro99,dor01}, and in particular their universal critical phenomena. In the Ising model at low temperatures, spins preferentially align with their neighbours, which is similar in spirit to the local maximisation behaviour of agents in the Naming Game. Here it has been found that Voter-type coarsening lies at a boundary between phases characterised the presence and absence of order. The phase diagram we have found here has rather similar characteristics, if one interprets the metastable state of the delayed-consensus phase as a disordered phase. In contrast to other works, our focus has been on statistics of the transit to an absorbing state in a finite system, as opposed to universal features of the stationary state \cite{oli93} or of coarsening and persistence \cite{dro99,dor01} displayed by infinite systems. It is not clear (apart, perhaps, from the overall timescales involved) that these contrasting properties are straightforwardly related.

An important question left open by our work concerns the effect of community structure on the consensus dynamics. The behaviour of the Voter Model in models with almost arbitrary structure is by now well understood, since in that case the essential contribution to the dynamics is captured by a single `collective coordinate' \cite{suc05,soo05,soo08,bax08a}. On the other hand, it is known that a system evolving by zero-temperature Glauber dynamics (which corresponds to an extreme form of maximising behaviour) on finite networks can become trapped in disordered metastable configurations \cite{cas05}. It would be interesting to examine the finite-temperature of the present model on more general network structures. One possibility here might be to try and expand around the analytically tractable Voter Model as a means to understand the crossover regime between the two phases.

\begin{table}
\begin{center}
 \begin{tabular}{l|c|l}
  Variant & Initial frequency & Source \\ \hline
  H-retention & 0.75 & p116 \\
  Weak vowel schwa & 0.32 & p117 \\
  Short front {\sc TRAP} vowel & 0.60 & p121 \\
  Diphthong shift & 0.75 & p121--2 \\
  Rounded {\sc LOT} vowel & 0.53 & p122 \\
  {\sc DANCE} vowel & 0.52 & p122 \\
  Fronted and lowered {\sc STRUT} & 0.34 & p136
 \end{tabular}
\end{center}
\caption{\label{tabtru} Numerical estimates for initial frequencies of conventionalised variants in the formation of the New Zealand English language dialect, taken from the indicated pages of \cite{tru04}.}
\end{table}

\begin{figure}
\begin{center}
 \includegraphics[width=0.75\linewidth]{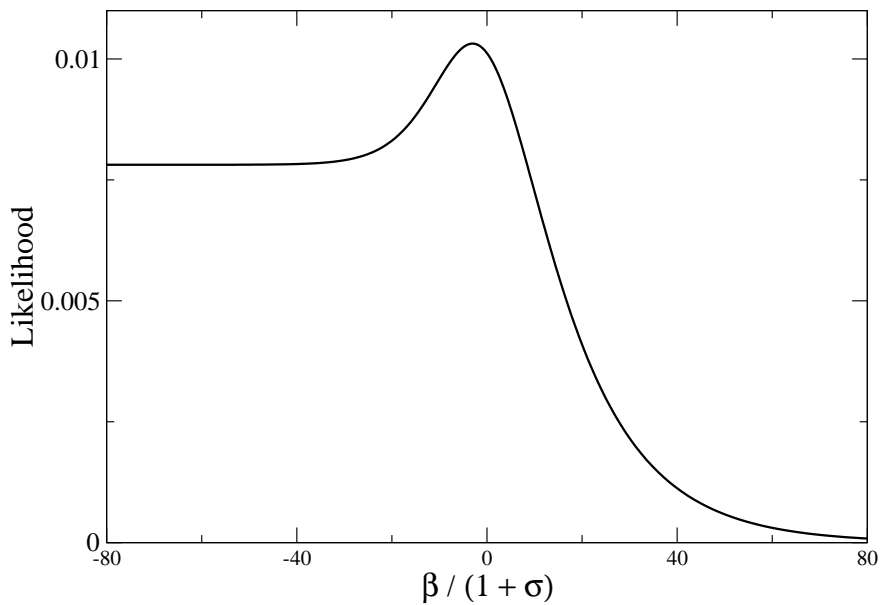}
\end{center}
 \caption{\label{NZlikelihood} Probability of observing the set of outcomes documented in the formation of the New Zealand English dialect \cite{gor04,tru04} within the hybrid model given the combination $\lambda=\frac{\beta}{1+\sigma}$ appearing in the consensus probability function $Q(x)$.}
\end{figure}

One may also legitimately ask what kind of parameter settings best describe a real human system. Here, relevant data are thin on the ground, although one pertinent set is provided by a thorough study of the emergence of the New Zealand English language dialect \cite{gor04,tru04}. This was formed through contact between different British and Irish dialects that arose through immigration to New Zealand in the mid 19th Century. In \cite{tru04}, seven features that initially exhibited variability are identified with estimates of the initial frequency of the surviving variant. These data are summarised in Table~\ref{tabtru}.  Assuming that the features evolved independently, we can ask how likely, given some combination of $b$ and $c$ in the model, consensus would have been reached on the set of variants with the initial frequencies $x_i$ specified in the table by calculating $\prod_i Q(x_i)$. We can then maximise this probability with respect to the model parameters to identify the instance of the hybrid model that best describes these empirical data.  For large $N$, $Q(x)$ depends in its central region on $b$ and $c$ through the parameter combination $\lambda = \beta/(1+\sigma)$ in both the rapid- and delayed-consensus phases. So, in fact, the best we can do is find the maximum likelihood value of $\lambda$. The likelihood as a function of $\lambda$ is plotted in Fig.~\ref{NZlikelihood}, from which we see that the data are best described by a value of $\lambda$ that is negative and of order unity. Assuming that $\beta$ is an increasing function of population size in structured as well as unstructured populations, and given that the number of New Zealand English speakers during the relevant historical period was over $10^5$, one may argue that if the community is well described by a single maximisation strategy, it is, in fact, a weak bias \emph{against} maximisation---perhaps as small as $10^{-5}$, if $\beta \propto N$ as it is in an unstructured community.  This observation may justify the use of the Utterance Selection Model as a means to critique theories for new-dialect formation in New Zealand \cite{bax08b}, although this should be cautioned by the fact that models allowing, for example, individual differences between speakers, and for their strategies to change over time (as suggested by the experimental study of \cite{hud05}), may fit the data better. These extensions to the model we leave as further possibilities for future study.

\section*{Acknowledgements}

The author thanks Bill Croft and Simon Kirby for linguistic insights, Martin Evans for comments on the manuscript and the RCUK for the support of an Academic Fellowship.

\appendix

\section{Symmetry relation for consensus time integral}

In this Appendix we show that the integral $I(x)$, defined through (\ref{Ix}), satisfies the relation (\ref{Isym}),
\begin{equation}
\label{limI}
\lim_{\beta\to\infty} \left[ I(x) + I(1-x) - I(1) \right] = 0
\end{equation}
in the range of $x$ that satisfies $\frac{1}{\beta} \ll x \ll \frac{1}{2}-\frac{1}{\sqrt\beta}$ as the limit is taken. We begin by noting that the integrand appearing in $I(1)$ is symmetric about $u=\frac{1}{2}$, and so we may write
\begin{equation}
\fl I(1) = \frac{\kappa\sqrt{\beta}}{2} \int_0^x {\rm d}u \frac{{\rm e}^{-\beta [V(u)-V(\frac{1}{2})]}}{b(u)} \left[ 1 - y(u)^2 \right] +  \frac{\kappa\sqrt{\beta}}{2} \int_0^{1-x} {\rm d}u \frac{{\rm e}^{-\beta [V(u)-V(\frac{1}{2})]}}{b(u)} \left[ 1 - y(u)^2 \right]
\end{equation}
where
\begin{equation}
 \kappa = \frac{\sqrt{\beta}}{2} \int_0^1 {\rm d}u {\rm e}^{\beta [V(u)-V(\frac{1}{2})]} \sim \frac{\sqrt{\pi (1+\sigma)}}{2}
\end{equation}
is a constant independent of $\beta$, $b(u) = u(1-u)[1+4\sigma u(1-u)]$ and $V(u) = \frac{1}{4\sigma} \ln[1+4\sigma u(1-u)]$.

We may then write
\begin{equation}
 \label{IJK}
 I(x) + I(1-x) - I(1) = \kappa\sqrt{\beta} \left[ J(x) + K(x) \right]
\end{equation}
where
\begin{eqnarray}
 J(x) &=& \int_0^x {\rm d}u \frac{{\rm e}^{-\beta [V(u)-V(\frac{1}{2})]}}{b(u)} \Bigg[ \frac{[y(x)-y(u)][1+y(u)]}{1+y(x)} - {}\\\nonumber && \qquad \frac{[y(x)+y(u)][1+y(u)]}{1-y(x)} - [1-y(u)^2] \Bigg] \\
 K(x) &=& - \int_{x}^{1-x} {\rm d}u \frac{{\rm e}^{-\beta [V(u)-V(\frac{1}{2})]}}{b(u)} \times \nonumber\\ && \qquad \Bigg[ \frac{[y(x)+y(u)][1+y(u)]}{1-y(x)} + \frac{1-y(u)^2}{2} \Bigg] \;.
\end{eqnarray}
To arrive at these expressions we have used the antisymmetry property $y(1-x)=-y(x)$.

Rearranging,
\begin{eqnarray}
 J(x) &=& - \frac{1+y(x)^2}{1-y(x)^2} \int_0^x {\rm d}u \frac{{\rm e}^{-\beta [V(u)-V(\frac{1}{2})]}}{b(u)} [1+y(u)]^2  \\
 K(x) &=& - \frac{1}{2} \frac{1+y(x)}{1-y(x)} \int_x^{1-x} {\rm d}u \frac{{\rm e}^{-\beta [V(u)-V(\frac{1}{2})]}}{b(u)} [1+y(u)]^2 \;.
\end{eqnarray}
In the integral for $J(x)$, we have that both $x$ and $u$ are much less than $\frac{1}{2} - \frac{1}{\sqrt{\beta}}$ and so we may use (\ref{yasympleft}) to approximate both $y(x)$ and $y(u)$. Bearing in mind also that $x \gg \frac{1}{\beta}$ we find to leading order
\begin{equation}
 J(x) \sim - \frac{2}{\kappa \sqrt{\beta}} \frac{{\rm e}^{-\beta V(x)}}{V'(x)} \int_0^x {\rm d}u \frac{{\rm e}^{-\beta V(u)}}{b(u)} \left[ \frac{{\rm e}^{\beta V(u)}}{V'(u)} -1 \right]^2
\end{equation}
This integral is dominated by its upper endpoint (the integrand vanishes linearly with $u$ as $u\to0$), and so we can drop the $1$ that appears in the square brackets. Then the leading behaviour of the integral is
\begin{equation}
 J(x) \sim - \frac{2}{\kappa \sqrt{\beta}} \frac{1}{V'(x)} \int_0^x {\rm d}u \frac{{\rm e}^{\beta [V(u)-V(x)]}}{V'(u)^2 b(u)} \;.
\end{equation}
Expanding around the endpoint $u=x$, we find that $J(x) \sim O(\beta^{-3/2})$.

For the $K(x)$ integral meanwhile, we find with a similar approximation for $y(x)$ that
\begin{equation}
 K(x) \sim - \frac{1}{2\kappa\sqrt{\beta}} \frac{1}{V'(x)} \int_x^{1-x} {\rm d}u \frac{{\rm e}^{-\beta[V(u)-V(x)]}}{b(u)} [1 + y(u)^2] \;.
\end{equation}
The dominant contributions to this integral come from the endpoints, and so by performing the same expansion as previously we find similarly that $K(x) \sim O(\beta^{-3/2})$. Hence inserting into (\ref{IJK}) we find that the limit in (\ref{limI}) is zero, as required.


\section*{References}

\begin{thebibliography}{10}
\providecommand{\url}[1]{\texttt{#1}}
\providecommand{\urlprefix}{URL }
\expandafter\ifx\csname urlstyle\endcsname\relax
  \providecommand{\doi}[1]{doi:\discretionary{}{}{}#1}\else
  \providecommand{\doi}{doi:\discretionary{}{}{}\begingroup
  \urlstyle{rm}\Url}\fi
\providecommand{\eprint}[2][]{\url{#2}}

\bibitem{kno04}
W.~Knospe, L.~Santen, A.~Schadschneider and M.~Schreckenberg (2004).
\newblock Empirical test for cellular automaton models of traffic flow.
\newblock \emph{Phys. Rev. E} \textbf{70} 016115.

\bibitem{hel07}
D.~Helbing, A.~Johansson and H.~Z. Al-Abideen (2007).
\newblock The dynamics of crowd disasters: an empirical study.
\newblock \emph{Phys. Rev. E} \textbf{75} 046109.

\bibitem{del07}
D.~{Delli Gatti}, E.~Gaffeo, M.~Allegati, G.~Guilioni, A.~Kirman, A.~Palestrini
  and A.~Russo (2007).
\newblock Complex dynamics and empirical evidence.
\newblock \emph{Information Sciences} \textbf{177} 1204.

\bibitem{cas07}
C.~Castellano, S.~Fortunato and V.~Loreto.
\newblock Statistical physics of social dynamics.
\newblock \eprint{arxiv:0710.3256}.

\bibitem{cav81}
L.~L. Cavalli-Sforza and M.~W. Feldman (1981).
\newblock \emph{Cultural Transmission and Evolution: A Quantitative Approach}
  (Princeton University Press, Princeton, NJ).

\bibitem{boy05}
R.~Boyd and P.~J. Richerson (2005).
\newblock \emph{The Origin and Evolution of Cultures} (Oxford University Press,
  Oxford).

\bibitem{pie03}
J.~B. Pierrehumbert (2003).
\newblock Phonetic diversity, statistical learning, and acquisition of
  phonology.
\newblock \emph{Lang. Speech} \textbf{46} 115.

\bibitem{cro00}
W.~Croft (2000).
\newblock \emph{Explaining language change: An evolutionary approach} (Longman,
  Harlow).

\bibitem{ewe04}
W.~J. Ewens (2004).
\newblock \emph{Mathematical Population Genetics: {I}: Theoretical
  Introduction} (Springer, New York).

\bibitem{lab01}
W.~Labov (2001).
\newblock \emph{Principles of lingustic change {II}: Social factors}
  (Blackwell, Oxford).

\bibitem{gor04}
E.~Gordon, L.~Campbell, J.~Hey, M.~MacLagan, A.~Sudbury and P.~Trudgill (2004).
\newblock \emph{New Zealand English: Its Origins and Evolution} (Cambridge
  University Press, Cambridge).

\bibitem{rog03}
E.~M. Rogers (2003).
\newblock \emph{Diffusion of innovations} (Free Press, New York), 5th edition.

\bibitem{cli73}
P.~Clifford and A.~Sudbury (1973).
\newblock A model for spatial conflict.
\newblock \emph{Biometrika} \textbf{60} 581.

\bibitem{bax06}
G.~J. Baxter, R.~A. Blythe, W.~Croft and A.~J. McKane (2006).
\newblock Utterance selection model of language change.
\newblock \emph{Phys. Rev. E.} \textbf{73} 046118.

\bibitem{oli93}
M.~J. {de Oliveira}, J.~F.~F. Mendes and M.~A. Santos (1993).
\newblock Non-equilibrium spin models with {Ising} universal behaviour.
\newblock \emph{J. Phys. A: Math. Gen.} \textbf{26} 2317.

\bibitem{dro99}
J.-M. Drouffe and C.~Godr\`eche (1999).
\newblock Phase ordering and perstitence in a class of stochastic processes
  interpolating between the ising and voter models.
\newblock \emph{J. Phys. A: Math. Gen.} \textbf{32} 249.

\bibitem{dor01}
I.~Dornic, H.~Chat\'e, J.~Chave and H.~Hinrichsen (2001).
\newblock Critical coarsening without surface tension: The universality class
  of the voter model.
\newblock \emph{Phys. Rev. Lett.} \textbf{87} 045701.

\bibitem{sch08}
F.~Schweitzer and L.~Behera (2008).
\newblock Nonlinear voter models: The transition from invasion to coexistence.
\newblock \eprint{arxiv:cond-mat/0307742v2}.

\bibitem{bar06}
A.~Baronchelli, M.~Felici, E.~Caglioti, V.~Loreto and L.~Steels (2006).
\newblock Sharp transition towards shared vocabularies in multi-agent systems.
\newblock \emph{J. Stat. Mech.: Theor. Exp.} \textbf{{}} P06014.

\bibitem{hud05}
C.~L. {Hudson Kam} and E.~L. Newport (2005).
\newblock Regularizing unpredictable variation: The roles of adult and child
  learners in language formation and change.
\newblock \emph{Lang. Learn. Devel.} \textbf{1} 151.

\bibitem{kir07}
S.~Kirby, M.~Dowman and T.~L. Griffiths (2007).
\newblock Innateness and culture in the evolution of language.
\newblock \emph{Proc. Nat. Acad. Sci.} \textbf{104} 5241.

\bibitem{bar07}
A.~Baronchelli, L.~{Dall'Asta}, A.~Barrat and V.~Loreto (2007).
\newblock Non-equilibrium phase transition in negotiation dynamics.
\newblock \emph{Phys. Rev. E} \textbf{76} 051102.

\bibitem{soo05}
V.~Sood and S.~Redner (2005).
\newblock Voter model on heterogeneous graphs.
\newblock \emph{Phy. Rev. Lett.} \textbf{94} 178701.

\bibitem{soo08}
V.~Sood, T.~Antal and S.~Redner (2008).
\newblock {Voter models on heterogeneous networks}.
\newblock \emph{Phys. Rev. E} \textbf{77} 41121.

\bibitem{bax08a}
G.~J. Baxter, R.~A. Blythe and A.~J. McKane (2008).
\newblock Fixation and consensus times on a network: A unified approach.
\newblock To appear in Phys. Rev. Lett.
\newblock \eprint{arxiv:0801.3083}.

\bibitem{bar08}
A.~Baronchelli, V.~Loreto and L.~Steels.
\newblock In-depth analysis of the {Naming Game} dynamics: the homogeneous
  mixing case.
\newblock To appear in Int. J. Mod. Phys.
\newblock \eprint{arxiv:0803.0398}.

\bibitem{tru04}
P.~Trudgill (2004).
\newblock \emph{New-dialect formation: The inevitability of colonial Englishes}
  (Edinburgh University Press, Edinburgh).

\bibitem{bly07}
R.~A. Blythe and A.~J. McKane (2007).
\newblock Stochastic models of evolution in genetics, ecology and linguistics.
\newblock \emph{J. Stat. Mech.:Theor. Exp.} \textbf{{}} P07018.

\bibitem{cas06}
X.~Castell\'o, V.~M. Egu{\'\i}luz and M.~{San Miguel} (2006).
\newblock {Ordering dynamics with two non-excluding options: Bilingualism in
  language competition}.
\newblock \emph{New J. Phys.} \textbf{8} 308.

\bibitem{kim69}
M.~Kimura and T.~Ohta (1969).
\newblock The average number of generations until fixation of a mutant gene in
  a finite population.
\newblock \emph{Genetics} \textbf{61} 763.

\bibitem{vaz04}
F.~Vazquez and S.~Redner (2004).
\newblock Ultimate fate of constrained voters.
\newblock \emph{J. Phys. A: Math. Gen.} \textbf{37} 8479.

\bibitem{GSL}
M.~Galassi, J.~Davies, J.~Theiler, B.~Gough, G.~Jungman, M.~Booth and F.~Rossi
  (2006).
\newblock \emph{GNU Scientific Library Reference Manual} (Network Theory,
  Bristol), 2nd edition.

\bibitem{red01}
S.~Redner (2001).
\newblock \emph{A Guide to First-Passage Processes} (Cambridge University
  Press, Cambridge).

\bibitem{ris89}
H.~Risken (1989).
\newblock \emph{The {Fokker-Planck} equation} (Springer-Verlag, Berlin).

\bibitem{gra00}
I.~S. Gradshteyn and I.~M. Ryzhik (2000).
\newblock \emph{Table of Integrals, Series and Products} (Academic Press,
  London).

\bibitem{all05}
R.~J. Allen, P.~B. Warren and P.~R. {ten Wolde} (2005).
\newblock Sampling rare switching events in biochemical networks.
\newblock \emph{Phys. Rev. Lett.} \textbf{94} 018104.

\bibitem{suc05}
K.~Suchecki, V.~M. Egu\'{\i}luz and M.~{San Miguel} (2005).
\newblock Conservation laws for the voter model in complex networks.
\newblock \emph{Europhys. Lett.} \textbf{69} 228.

\bibitem{cas05}
C.~Castellano, V.~Loreto, A.~Barrat, F.~Cecconi and D.~Parisi (2005).
\newblock Comparison of voter and {Glauber} ordering dynamics on networks.
\newblock \emph{Phys. Rev. E} \textbf{71} 066107.

\bibitem{bax08b}
G.~J. Baxter, R.~A. Blythe, W.~Croft and A.~J. McKane.
\newblock Modeling language change: An evaluation of {Trudgill's} theory of the
  emergence of {New Zealand English}.
\newblock Submitted to Language Variation and Change.

\end{thebibliography}
\end{document}